\documentclass[12pt,preprint,nofootinbib,noshowkeys,noshowpacs,longbibliography,
superscriptaddress,tightenlines,floatfix]{revtex4-2}
\usepackage{amsmath,amsfonts,amsthm,amssymb,bbm}
\usepackage{graphics,graphicx}
\usepackage{color}
\usepackage{bm}
\definecolor{darkblue}{RGB}{0,0,196}
\definecolor{darkgreen}{RGB}{0,120,0}

\usepackage{stackengine}

\renewcommand\S{\mathcal S}

\renewcommand\P{\mathcal P}
\newcommand\N{\mathcal N}
\usepackage[inline]{enumitem}
\def\half{\frac{1}{2}}

\def\llangle{\left\langle}
\def\rrangle{\right\rangle}
\newcommand{\nn}{\nonumber}

\def\muh{\hat\mu}
\def\vh{\hat v}
\def\Eq#1{eq.~(\ref{#1})}
\def\Eqs#1{eqs.~(\ref{#1})}
\def\eq#1{(\ref{#1})}
\def\app#1{App.~\ref{#1}}
\def\Fig#1{Fig.~\ref{#1}}

\def\Sect#1{Section~\ref{#1}}

\def\beq{\begin{equation}}
\def\eeq{\end{equation}}
\def\st{\begin{equation}}
\def\stp{\end{equation}}
\def\ba{\begin{eqnarray}}
\def\ea{\end{eqnarray}}   
\def\dd{{\rm d}}
\newcommand\ph{\phantom{\nu}}
\usepackage[colorlinks=true,linktocpage=true,linkcolor=darkblue,citecolor=red,urlcolor=darkblue]{hyperref}

\def\x{{\bf x}}
\def\p{{\bf p}}
\def\DLF{{\rm \scriptscriptstyle D,LF}} 
\def\LF{{\rm \scriptscriptstyle LF}} 

\def\clp{{\mathcal C}_p}

\begin{document}

\preprint{}
 
    \title{The stochastic relativistic advection diffusion equation from the Metropolis algorithm}
    \author{G\"ok\c ce Ba\c sar}
    \email{gbasar@unc.edu}
	\affiliation{Department of Physics and Astronomy, University of North Carolina, Chapel Hill, North Carolina 27599, USA}
    \author{Jay Bhambure}
    \email{jay.bhambure@stonybrook.edu}	
    \affiliation{Center for Nuclear Theory, Department of Physics and Astronomy, Stony Brook University, Stony Brook, New York, 11794-3800, USA}
    \author{Rajeev Singh}
    \email{rajeevofficial24@gmail.com}
    \affiliation{Center for Nuclear Theory, Department of Physics and Astronomy, Stony Brook University, Stony Brook, New York, 11794-3800, USA}
    \affiliation{Department of Modern Physics, University of Science and Technology of China, Hefei, Anhui 230026, China}
    \author{Derek Teaney}
    \email{derek.teaney@stonybrook.edu}
    \affiliation{Center for Nuclear Theory, Department of Physics and Astronomy, Stony Brook University, Stony Brook, New York, 11794-3800, USA}
	\date{\today} 
	\bigskip
\begin{abstract}
  We study an approach to simulating the stochastic relativistic advection-diffusion equation based on the Metropolis algorithm. We show  that the  dissipative dynamics of the boosted fluctuating fluid can be simulated by making random transfers of charge between fluid cells, interspersed with ideal hydrodynamic time steps. The random charge transfers are accepted or rejected in a Metropolis step using the entropy as a statistical weight.  This procedure reproduces the expected stress of dissipative relativistic hydrodynamics in a specific (and non-covariant) hydrodynamic frame known as the \emph{density frame}.
  Numerical results, both with and without noise, are presented and compared to relativistic kinetics and analytical expectations.  An all order resummation of the density frame gradient expansion reproduces the covariant dynamics in  a specific model.
  In contrast to all other numerical approaches to relativistic dissipative fluids,
  the dissipative fluid formalism presented here is strictly first order in gradients and has no non-hydrodynamic modes.
  The physical naturalness and simplicity of the Metropolis algorithm, together with its convergence properties, make it a promising tool for simulating stochastic relativistic fluids in heavy ion collisions and for critical phenomena in the relativistic domain.
\end{abstract}
     
\date{\today}
	
\maketitle
\newpage

\section{Introduction}
\label{sec:intro}

%

\subsection{Physical motivation}
Nuclear collisions at high energy exhibit remarkable collective flows, which
have been analyzed with considerable success using relativistic hydrodynamics~\cite{Heinz:2013th}.
For large nuclei, ideal hydrodynamics provides a reasonable description of the
observed flows. Viscous corrections are then incorporated by simulating  
(a version of) the relativistic Navier-Stokes equations, improving the description of the data and clarifying the theoretical consistency of the simulations.
These simulations fit the shear viscosity  to entropy ratio of QCD
around the crossover temperature.  Current Bayesian fits give $\eta/s \simeq
2\, \hbar/4\pi k_B$~\cite{Bernhard:2019bmu,PhysRevLett.126.202301,PhysRevC.103.054904,Heffernan:2023kpm}, which  indicates that the medium is remarkably strongly coupled, with relaxation rates of the order $\sim k_BT/\hbar$. 

The hydrodynamic description of heavy ion collisions can be tested by examining
the collisions of light nuclei, such as  ${\rm d+Au}$ and ${\rm He}^3+{\rm Au}$
at the Relativistic Heavy Ion Collider (RHIC) and proton-nucleus collision at the Large Hadron Collider (LHC).
Remarkably, these events also exhibit correlations which are indicative of the collective flow~\cite{*[{See for example Sec. 3.2.2 and Sect 3.2.3 in: }][{.}] Arslandok:2023utm}.  However, it must be emphasized, that the
hydrodynamic description of these events is breaking down.  This is in  part
because (a suitably defined) mean free path $\ell_{\rm mfp}$ has become
comparable to the system size,  and in part because the total number of
particles produced in these events $N_{\rm ch}$ is becoming small, which leads to
large fluctuations. One of the motivations for the current paper is to analyze
thermal fluctuations in relativistic dissipative systems, with the ultimate
goal of describing small colliding systems.

The current manuscript is also motivated by two classical $2^{\rm nd}$ order phase transitions in QCD, which may be an observable in heavy ion collisions.  In both of these transitions
incorporating thermal fluctuations into the hydrodynamic
description is essential to describing the underlying physics in the critical region.
The first phase transition is the $O(4)$ chiral transition of QCD, which is a
$2^{\rm nd}$ order phase transition in the limit of two massless quark flavors~\cite{Pisarski:1983ms,Rajagopal:1992qz}. There is
a strong motivation from lattice QCD to look for signatures of the $O(4)$ transition in the
heavy ion data at the LHC~\cite{HotQCD:2019xnw,Kaczmarek:2020sif}.
At lower temperatures and higher baryon
density, strong theoretical arguments suggest that there should
be a Ising critical point in the $(T, \mu_B)$ plane~\cite{Stephanov:2004wx}, with $T$ and $\mu_B$ being the temperature and the baryon chemical potential, respectively. Currently at RHIC, there is an ongoing search for the Ising critical point where the beam energy is scanned in an effort to tune the  baryon chemical potential to the critical region of the phase diagram~\cite{Du:2024wjm}.
\subsection{The Metropolis algorithm for relativistic hydrodynamic fluctuations}

We are also motivated to study thermal fluctuations in relativistic fluids by algorithmic developments and the mathematical structure of stationary stochastic processes.
In statistical mechanics, the  Metropolis algorithm is used to generate  field configurations of a field $\phi$  from a known probability distribution, $P[\phi] \propto e^{S[\phi]}$,  where $S[\phi]$ is the entropy~\cite{Landau_Binder_2014}.  
In the algorithm a proposal is made for a change in the fields,  $\phi \rightarrow \phi + \Delta \phi$.  This proposal is  accepted or rejected according to the magnitude and sign of the change in the entropy, $\Delta S \equiv  S[\phi + \delta \phi] - S[\phi]$.  The Metropolis steps are guaranteed to converge to the required equilibrium distribution.  
Recently,  in the context of simulating the $O(4)$ critical point, we simulated the stochastic diffusion of a conserved charge coupled to the order parameter using a variant of the Metropolis algorithm~\cite{Florio:2021jlx}. A proposal is made for the  transfer of charge between the fluid cells, and this proposal is then accepted or rejected based on the change in the entropy of the system.
For small enough time steps  the Metropolis updates naturally reproduce the Langevin dynamics of the diffusion equation.  The equivalence of the Metropolis and Langevin dynamics for small time steps $\Delta t$ has been repeatedly observed over the years~\cite{OldMCMC,Moore:1998zk}.

The advantages of a Metropolis based approach is that detailed balance and the Fluctuation-Dissipation-Theorem (FDT) are automatically preserved,  independently of $\Delta t$, which  guarantees that Markov-chain will equilibrate to a specific action. In non-linear theories this simplifies the renormalization of the theory and clarifies  discretization ambiguities that arise in non-linear Langevin equations~\cite{Arnold:1999uza,Gao:2018bxz}. For fluids at rest, the Metropolis algorithm has been used to implement the non-linear Langevin dynamics in a number of challenging applications, leading to the study of sphaleron transitions of hot QCD~\cite{Moore:1998zk},  the real time dynamics of $O(4)$ critical point in QCD~\cite{Florio:2021jlx,Florio:2023kmy}, and the dynamics of Model B and Model H~\cite{Chattopadhyay:2023jfm,Chattopadhyay:2024jlh} in the Halperin-Hohenberg classification of dynamical critical phenomena.  



Our principal task in this and a companion paper is to generalize the Metropolis approach to relativistic fluids in general coordinates,  where the (somewhat complicated) form of the dissipative stress should arise naturally from the accept/reject steps of the Metropolis algorithm.  As a first step, in this paper we will consider the diffusion of charge in a relativistic fluid.

\subsection{Causality, second order hydrodynamics, and the Metropolis algorithm }

To understand the issues that arise with relativity, consider the relativistic
advection-diffusion equation in flat spacetime in the Landau-Lifshitz frame, momentarily neglecting the stochastic noise for simplicity.
We are considering  a fluid moving  with three velocity $v^i$ in the lab frame  and following the diffusion of a dilute conserved charge within the fluid,  $\partial_{\mu} J^{\mu}=0$.  The four velocity is $u^{\mu}=(\gamma, \gamma {\bf v})$ and the local charge density in the rest frame of the fluid is $n_\LF = -u_{\mu} J^{\mu}$. Landau and Lifshitz define a hydrodynamic frame where  the chemical potential  is given by  the value of $n$ and ideal equation of state to all orders in the gradient expansions, i.e. $n_\LF=\chi \, \mu_\LF$ where $\chi$ is the charge susceptibility~\cite{landau2013fluid}.  



The problem with the covariant Landau-Lifshitz approach is that the  diffusive current,  which was spatial in the rest frame of the fluid $j_D^i = - D \partial^i n$, involves time derivatives in an arbitrary frame. This leads to equations which are second order in time, which in turn leads to runaway solutions and other pathological behavior~\cite{Hiscock:1985zz,Gavassino:2021owo}.  One way to correct this pathology is to promote the diffusive current to an additional dynamical field  which relaxes on collisional timescale to the expected form.  This procedure results in Maxwell-Catteneo or Israel Stewart type equations.

There is merit to the Israel-Stewart approach -- it provides an effective way to realize the dynamics of the relativistic diffusion equation and the relativistic Navier Stokes equations more generally. Indeed, 
almost all large scale simulations of flow in heavy ion collisions are based on this approach.
 However, the Israel-Stewart formulation involves fast variables whose physical significance should be questioned.  Indeed, there have been many reformulations of viscous hydrodynamics in the relativistic domain~\cite{Israel:1976tn,Israel:1979wp,Geroch:1990bw,OTTINGER1998433,Bemfica:2017wps,Kovtun:2019hdm}, and  each of these reformulations involve some additional variables (or non-hydrodynamic modes) which relax quickly to the form constrained by ``first-order'' hydrodynamics~\cite{Lindblom:1995gp}. 
A kind of theorem has emerged, which states that it is impossible to construct 
a causal and stable relativistic theory of hydrodynamics without incorporating non-hydrodynamic modes~\cite{Heller:2022ejw,GAVASSINO2023137854,Gavassino:2023mad}.  

In this paper we will investigate an alternative to the Israel-Stewart approach 
developed to describe fluids without an underlying boost symmetry~\cite{Novak:2019wqg,deBoer:2020xlc,Armas:2020mpr}. In particular, we found the ``density frame'' discussed by Armas and Jain a clarifying formalism when implementing Metropolis updates~\cite{Armas:2020mpr}.  The density frame has no non-hydrodynamic modes and no additional parameters compared to the Landau theory of first order hydrodynamics, at the price of not being fully boost invariant.
In hydrodynamics without boosts the constitutive relations are written down for setups where the underlying interactions are not Lorentz invariant and thus there is a preferred ``lab" frame\footnote{For a physical example, consider a fluid flowing over a table and study the diffusion of charge in this background fluid flow. The table sets a preferred lab frame.}.
If the  microscopic interactions happen to be Lorentz invariant, the additional boost symmetry imposes relations between the coefficients of the  gradient expansion, which is an expansion in lab-frame spatial derivatives.
Indeed, the equations of motion in the density frame follow from the Landau ones if the ideal equations  are  used to rewrite lab-frame time derivatives appearing in the dissipative strains as  spatial derivatives. With this rewrite the equations are strictly first order in time and are stable. 

The procedure amounts to a non-covariant choice of hydrodynamic frame where the relation between the chemical potential and the lab frame charge per volume $J^0$  is given by ideal hydrodynamics at all orders in the gradient expansion, i.e. $J^0 = \chi \mu u^0$.
The frame choice and the resulting equations of motion in the density frame are
not invariant under Lorentz transformations; but, they are invariant under
Lorentz transformations followed by a change of hydrodynamic frame,  which
reparametrizes the hydrodynamic fields. The results obtained in different
Lorentz frames will vary, but the variation is beyond the accuracy of the
diffusion equation.  Different choices of Lorentz invariant hydrodynamic frames, such as the Landau, Eckart or  BDNK\footnote{The acronym is short for Bemfica, Disconzi, and Noronha~\cite{Bemfica:2017wps} and Kovtun~\cite{Kovtun:2019hdm}. These authors studied a class of Lorentz invariant frames where the temperature is related to the Landau choice up to derivatives.} choices, will also give different (if Lorentz invariant) results to this accuracy.  The density frame approach was known to ``work'' in simple cases,  but the work on hydrodynamics without boosts formalized the procedure. Finally, Armas and Jain made an important connection of this approach to modern treatments of hydrodynamics~\cite{Crossley:2015evo,Glorioso:2017fpd} where the conserved charge and the corresponding canonical conjugates (for instance a $U(1)$ charge $Q$ and the associated phase $\varphi$) play a dual role~\cite{Armas:2020mpr}.  The numerical utility of the symplectic structure inherent in this duality remains to be fully exploited.


The structure of the current paper is as follows. Section~\ref{sec:advect_diffuse_eqn} discusses the relativistic advection-diffusion equation in the Landau and density frames, and derives the density frame constitutive relation from covariant kinetic theory. Section~\ref{sec:comparison_kinetics} compares the density frame relativistic advection-diffusion equation with relativistic kinetics. As discussed above, the density frame is not Lorentz invariant,  but it is invariant under Lorentz transformations followed by a reparametrization of the hydrodynamic variables. In a specific test problem discussed in \Sect{sec:comparison_kinetics}, we show that, in the regime of validity of hydrodynamics, the deviations of the density frame from a underlying covariant microscopic theory are controllably small,  even for highly boosted fluids.  We also  study the convergence of the gradient expansion of the density frame, making  connections with Lorentz covariant approaches.  Stochastic dynamics in the density frame is studied in \Sect{sec:stochastic_dynamics}. 
Since the hydrodynamics equations are defined  using a  given foliation of space-time, the Metropolis updates discussed above are very natural and easy to implement.  One simply makes random transfers of charge  in between ideal advective steps. These charge transfers are then accepted or rejected using 
 the entropy defined on the spatial slice as the statistical weight.  We present a first study of numerical correlation functions from the Metropolis algorithm for an equilibrated boosted fluid in \Sect{sec:stochastic_dynamics}. Although we have used the Metropolis algorithm for the relativistic hydrodynamics in the density frame, it should be useful for other approaches to stochastic relativistic hydrodynamics, e.g. approaches based on BDNK~\cite{Mullins:2023tjg,Gavassino:2024vyu} or Israel-Stewart~\cite{Singh:2018dpk}.
Finally,  in \Sect{sec:discussion} we conclude with a short discussion of the next steps.


\section{The advection diffusion equation}
\label{sec:advect_diffuse_eqn}
\subsection{Setup and first order hydrodynamics in the Landau frame}

We are considering the advection and diffusion of a charge in fluid  moving at relativistic speeds in flat spacetime, $\eta_{\mu\nu} = (-,+,+,+)$. 
The charge density is low and the temperature and flow velocity $u^{\mu} = (\gamma, \gamma {\bf v})$  may be considered fixed. (The generalization of this problem to a time dependent background fluid is discussed in \app{app:spacetimedepend}.) In first order hydrodynamics in the Landau frame the conserved current obeys~\cite{landau2013fluid}
\st
 \partial_{\mu} J^{\mu} = 0\,,  \qquad  J^{\mu}  \equiv n_{\LF} u^{\mu}   + j_{\DLF}^{\mu}   \, , 
\stp
where the first term in $J^{\mu}$ is ideal advection and  the second term is the diffusive correction, expressed as
\st
  j_{\DLF}^{\mu}  \equiv - T\sigma \Delta^{\mu\nu} \partial_{\nu} \muh_{\LF} \, .
\stp
Here  $T$ is the temperature, $\sigma$ is the conductivity, $\muh_{\LF} \equiv \mu/T$ is the scaled 
chemical potential, thermodynamically conjugate to charge density $n_\LF$.  $\Delta^{\mu\nu}$ is the spatial projector 
\st
        \Delta^{\mu\nu} = \eta^{\mu\nu} + u^{\mu} u^{\nu}  \, , 
\stp
and satisfies $\Delta^{\mu\rho} \Delta_{\rho\nu} =  \Delta^{\mu}_{\ph \nu}$.  
Since the density is low, $n_{\LF} = \chi\mu_{\LF}$ where $\chi$ is the (temperature dependent) susceptibility. 

In the next subsections we will briefly review the density frame pointing out the differences with the Landau frame. To keep the presentation self contained and pedagogical, sections~\ref{subsec:thermodynamics} and \ref{subsec:advect_diffuse_density_frame} review a small portion of \cite{Armas:2020mpr} with a focus on the diffusion equation. Section~\ref{subsec:densityframe_kinetics} describes how the density frame constitutive relation arises in relativistic kinetic theory.

\subsection{Thermodynamics of a boosted fluid}
\label{subsec:thermodynamics}
Consider a portion of a fluid in perfect global equilibrium
within a  lab frame measurement volume,  $V_0  \equiv \int \dd \Sigma_0 $.
The entropy, energy-momentum, and charge on this slice are 
\st
     \mathcal S = V_0 S\,,   \qquad  \mathcal P_{\mu} \equiv V_0 T^{0}_{\ph \mu}\,,    \qquad  \mathcal N = V_0 J^0\,.
\stp
We will notate the charge density with $N \equiv J^0$, and the temporal components of the energy momentum tensor with $(E, M^i) \equiv (T^{00}, T^{0i} )$, reserving the calligraphic symbols $\mathcal S$, $\P_\mu$ and  $\N$ for the charges,  as opposed to the charge densities, $S, E, M, N$.
When a boost symmetry is ultimately adopted, the entropy will take the form  $S = s u^0$ where
$s$ is a Lorentz scalar.

Given the conserved charges $\P_{\mu}$ and $\N$,  
the temperature, velocity and chemical potential can be determined from 
the micro-canonical equation of state $S(\P, \N, V_0)$
\begin{align}
\label{eq:thermo1}
\dd \mathcal S =& -\beta^{\mu} \dd\P_{\mu} - \muh \, \dd \N  + p \beta^0  \dd V_0 \,,  \\
      =&  -\beta^{0} \dd \P_0  - \beta^0 v^i \dd \P_i - \muh \, \dd \N  + p \, \beta^0 \dd V_0\,,
 \end{align}
where $\muh \equiv \mu/T$  and $v^i \equiv \beta^i/\beta^0$ is the fluid velocity.
The Gibbs-Duhem relation follows from the extensivity of the system
\st
\label{eq:thermo2}
  S = -\beta^{\mu} T^{0}_{\ph \mu}  - \muh J^0   + p \, \beta^0 \,.
\stp
Then for small charge densities the entropy density as a function of $N$ takes the form
\st
\label{eq:Sdensity}
 S(N) = \underbrace{ S_1(E,M)  }_{ \mbox{const} } - \frac{\beta^0}{2 \chi^{00}} N^2 \,.
\stp
So far we have not used boost symmetry, and the  parameters, such as $\beta^0$, $\muh$  and 
$\chi^{00}$ are functions of $E$ and $M$. After imposing boost symmetry in the next paragraph, $\beta^0$ will be a Lorentz scalar $\beta$ times $u^0$, and $\chi^{00}$ will be a Lorentz scalar $\chi(\beta)$ times $u^0 u^0$
\st
 \beta^0 = \beta u^0\,, \qquad  \chi^{00} = \chi(\beta) u^0 u^0\, , 
\stp
justifying the notation a posteriori.



When the fluid has an underlying Lorentz symmetry the dependence on the velocity is determined by the symmetry. Before imposing the symmetry, the conservation laws take the form
\begin{align}
  \partial_t N + \partial_i J^i =&0 \,, \nonumber \\
  \partial_t E + \partial_i T^{i0} =&0 \,, \nonumber \\
  \partial_t M^j + \partial_i T^{ij} =&0 \,. 
  \label{eq:ideal_hydro}
\end{align}
At zeroth order in the gradient expansion (ideal hydrodynamics) the three current $J^i$, the energy flux $T^{i0}$, and  the spatial stress tensor $T^{ij}$  are algebraically related to the 
conserved
charges $E$, $M^i$, and $N$. 
Using the conservation laws, the thermodynamic relations (\Eqs{eq:thermo1} and \eqref{eq:thermo2}),    the symmetry of the stress tensor imposed by Lorentz invariance $T^{i0} = M^i$, and  requiring that  $\partial_t S + \partial_i S^i$ be non-negative, fixes the form of fluxes $J^i$, $M^i$, $T^{ij}$  and entropy current $S^i$ to the form of ideal hydrodynamics~\cite{Armas:2020mpr}.  In particular,  the charges and entropy  are parametrized by $\beta^\mu\equiv \beta u^{\mu}$ and the chemical potential $\muh$
\st
\label{eq:ideal_hydro_relations}
   E = (e + p) (u^0)^2 - p , \qquad M^i = (e+p) u^0 u^i , \qquad
  N = n u^0  , \qquad S = s u^0 .
\stp
Here $e$, $p$, $n$ and $s$ are scalar functions of $\beta \equiv \sqrt{- \beta^{\mu} \beta_\mu}$ and $\muh$ as given by the equilibrium equation of state.  
At small chemical potentials, the local charge density and entropy takes the form 
\st
\label{eq:slorentz}
        n = \chi \, \mu \,, \qquad  s(\beta, \mu) = s_1(\beta) - \tfrac{1}{2} \beta \chi \mu^2 \,.
\stp
where $\chi(\beta)$ is a function of temperature. Comparing the definitions in \Eq{eq:Sdensity} and \Eq{eq:slorentz}, we find
$\chi^{00}/\beta^0 = T\chi u^0$  as claimed above. 

In the density frame the familiar relations of ideal hydrodynamics, eqs.~\eqref{eq:ideal_hydro_relations}, serve to define the  temperature, chemical potential, and flow velocity in terms of the lab frame charges, $E$, $M^i$ and $N$,  at every order in the gradient expansion, i.e. the relation between the charges  $E, M^i, N$ and their conjugates do not receive viscous corrections.  In particular, the chemical potential in the density frame is  defined at all orders in the gradient expansion as
\st
 \mu = \frac{J^0}{\chi u^0}\,.
\stp
This definition should be contrasted with the chemical potential in the Lorentz invariant Landau frame where 
\st
 \mu_{\LF} = - \frac{u_{\mu} J^{\mu}}{\chi} \, .
\stp
With the density frame definition a fiducial observer needs to count the charge in a given measurement volume in order to determine the chemical potential.
With the  Landau frame definition,  the three current $J^{i}$  also needs to be measured. Thus, the Landau frame  involves counting the charges at different times in order to define the chemical potential.

\subsection{The advection diffusion equation in the density frame}
\label{subsec:advect_diffuse_density_frame}

Here we will derive the density frame equations of motion by first considering 
fluids without a boost symmetry, and then specializing the equations to Lorentz 
covariant fluids. An example of a two dimensional non-Lorentz invariant fluid is a fluid 
flowing over a flat surface at relativistic speeds. The diffusion of a charge in 
this fluid depends on the speed of the fluid relative to the surface.

The advection-diffusion equation in the density frame consists of the conservation 
law
\st
 \partial_t N + \partial_i J^i = 0 \,,
\stp
together with a constitutive relation  for the diffusive current $J_D^i$
\st
 J^i \equiv N v^i +   J_D^i \,.
\stp
The diffusive current is expanded in spatial gradients of the conserved charge, or its thermodynamic conjugate $\muh$.  The most general form of $J_{D}^i$ at first order in gradients of $\muh$ is 
\st
\label{DFframe1}
 J_D^i = - \frac{\sigma_\parallel(\beta^0, v)}{\beta^0} \,  \vh^i \vh^j \partial_{j}\muh  - \frac{\sigma_\perp(\beta^0, v)}{\beta^0} \left(\delta^{ij} - \vh^i \vh^j \right) \partial_j  \muh \,,
\stp
where $\muh \equiv \partial S/\partial N = \beta^0 N/\chi^{00}$ and $\vh^i = v^i/|v|$ is a flow unit vector.
The first and second terms on the right hand side of \Eq{DFframe1} capture the diffusion parallel and perpendicular  to the fluid motion respectively.
Demanding that entropy production be positive leads to the requirement that $\sigma_{\parallel}(\beta^0,v) > 0$ and $\sigma_\perp(\beta^0, v) > 0$, but no further constraints can be derived on general grounds.

Lorentz invariant fluids can be treated as a special case of \Eq{DFframe1}.  Indeed, 
the boost symmetry determines the  dependence of  $\sigma_\parallel$ and $\sigma_\perp$  on the velocity.
The easiest way to derive this relation is to return momentarily  to the Landau frame.
Comparison to the density frame form gives
\st
  N = n_{\LF} u^0 + j_\DLF^0 \,,
\stp
and 
\st
\label{eq:jdi}
J_D^i = J^i  - N v^i = (\Delta^i_{\ph \alpha} - v^i \Delta^{0}_{\ph \alpha}) j_\DLF^{\alpha} \,.
\stp
We  now use the lowest order equation  of motion\footnote{When the fluid velocity
is not constant this expression receives additional corrections proportional to 
derivatives of velocity. The generalized constitutive relation is given in \app{app:spacetimedepend}.},
\st
  \partial_t\muh  \simeq  -v^j\partial_j \muh \,,
\stp
to approximate the Landau frame  expression for the diffusive current 
\st
\label{eq:jdlf}
  j_{\DLF}^\alpha \simeq -T\sigma \left(\Delta^{\alpha j } - \Delta^{\alpha 0} v^j \right) \partial_j\muh \,.
\stp
Substituting \Eq{eq:jdlf} into \Eq{eq:jdi} gives the current in the density frame
\st
\label{eq:JDi}
  J_D^{i} = -T \, {\sigma}^{ij} \, \partial_j \muh \,,
\stp
where we have defined a frequently occurring matrix 
\begin{align}
  T\sigma^{ij} =& T\sigma\left(\Delta^{i}_{\ph \alpha} - v^i \Delta^{0}_{\ph \alpha} \right) \left( \Delta^{j}_{\ph \beta} - v^j \Delta^{0}_{\ph \beta }  \right) \,  \Delta^{\alpha\beta}  \, , \\ 
   =&  T\sigma \left(\delta^{ij} - v^i v^j\right) \,. 
   \end{align}
   A generalization of the constitutive relation in \eqref{eq:jdlf} for  fluids depending on space and time is given \app{app:spacetimedepend}. 

Comparison  with the general form in \Eq{DFframe1}  shows that
\st
 \frac{\sigma_\parallel(\beta^0, v)}{\beta^0} =   \frac{T\sigma(\beta)}{\gamma^2 }\,, \qquad \qquad   \frac{\sigma_\perp(\beta^0, v)}{\beta^0} = T\sigma(\beta) \,.
\stp
In summary, in the density frame the equation of motion is 
\st
  \partial_t N  + \partial_i (N v^i)  = \partial_{i}\left(T \sigma^{ij} \partial_j  \muh\right)\,,
\stp
and when $\muh$ is written in terms of the charge $N = \chi \mu u^0$,  we arrive at an advection-diffusion equation 
\st
  \partial_t N  + \partial_i (N v^i)  = \partial_{i}\left(D^{ij} \partial_j N\right) \,,
\stp
with a diffusion matrix
\st
\label{eq:Dmatrix}
  D^{ij}  =\frac{D}{\gamma} \left(\delta^{ij}  - v^{i} v^j \right) \,.
\stp
Here $D=\sigma/\chi$ is the scalar diffusion coefficient of the Landau frame. Apart from 
the tensor structure  the equation is numerically similar to the non-relativistic advection-diffusion equation and can be solved by familiar numerical methods\footnote{In our numerical tests without noise we used an IMEX scheme with a standard advection step and a  Crank-Nicholson like implicit step using the {\tt PETSc} library~\cite{petsc-web-page}. }.

The $\gamma$ factors in the diffusion matrix can be easily understood physically.  The diffusion coefficient has units of distance squared per time.   The rate of transverse diffusion is suppressed relative to a fluid at rest  by one factor of $\gamma$ due to time dilation. The rate of longitudinal diffusion is suppressed by three factors of $\gamma$ due to time dilation and length contraction, i.e. each spatial step in the random walk  is length contracted by $\gamma$ and the steps add in square.

\subsection{The density frame from relativistic kinetics }
\label{subsec:densityframe_kinetics}
In this subsection we will briefly describe how the density frame constitutive relation arises naturally in relativistic kinetic theory. 
Specifically, we will show how the conductivity matrix $T\sigma^{ij}$ follows from  covariant 
kinetics  and find how this matrix is determined by the particles through the first viscous correction $\delta f$ to the phase-space distribution function. 
For simplicity, we will  assume a relaxation time approximation and consider single species
of classical relativistic particles, which carry the charge of the system  
\st
  p^{\mu} \partial_{\mu} f =  -\clp \, \delta f \,.
\stp
Here $\clp$ is a momentum dependent  parameter controlling the collision rate 
in the rest frame of the medium.

In global equilibrium the phase space distribution function is characterized by a constant chemical potential, temperature, and flow velocity.
If the density of the charged particles depends slowly on space and time then the parameter $\muh(t,{\bf x})$  is no longer a constant  but reflects this dependence
\st
f_0(t,\x,\p) =  e^{\muh(t,\x)} e^{\beta^{\mu} p_{\mu}} \,.
\stp
In the density frame $\muh(t,{\bf x})$ is adjusted to reproduce the charge density in the lab frame $J^0$, while in the Landau frame $\muh(t,{\bf x})$ is adjusted to reproduce the charge density in the rest frame, $n(t,{\bf x}) = -u_{\mu} J^{\mu}$. 
These two definitions of the chemical potential agree when gradients are 
neglected,  and in this case $f_{\rm eq} (t,{\bf x}, \p)$ is a solution to the Boltzmann equation.
$\muh(t,{\bf x})$ obeys the equations of ideal advection equation  at lowest order
\st
 u^{\mu} \partial_{\mu} \muh \simeq 0  \,.
 \label{eq:lowestEOM}
\stp

Following a standard procedure to find the first viscous correction~\cite{Teaney:2009qa}, we parameterize $f = f_0 + \delta f(t, \x, \p)$
and solve for $\delta f$ order by order in the gradients
\st
   f_0 \,  p^{\mu} \partial_{\mu} \muh  = - \clp \, \delta f \,.
   \label{eq:gradients}
\stp
In the Landau frame one decomposes the gradient into its temporal and spatial components as
\st
  \partial_{\mu} \muh =  -u_{\mu} u^{\alpha}\partial_\alpha \muh  + \Delta_{\mu}^{\ph \alpha}\,  \partial_\alpha \muh \, .
  \label{eq:gradients_components}
\stp
We neglect the temporal term in Eq.~\eqref{eq:gradients_components} exploiting the lowest order equations of motion, Eq.~\eqref{eq:lowestEOM}. Then we substitute into Eq.~\eqref{eq:gradients}, 
which leads to the familiar form of the first viscous correction in the Landau frame
\st
   \delta f_\LF = -\clp^{-1} \, f_0\,  p^{\alpha}  \nabla_\alpha \muh_\LF \, ,
\stp
where $\nabla_\alpha = \Delta_{\alpha}^{\ph \mu} \partial_\mu$.
Evaluating the diffusive current
\st
   j^{\mu}_{\DLF} = \int_p \frac{d^3p}{(2\pi)^3}  \, \frac{p^\mu}{p^0} \delta f_\LF    \,,
\stp
yields the expected form of the current in the Landau frame
\st
 j_\DLF^{\mu} = -T\sigma \Delta^{\mu \nu} \partial_\nu \muh_\LF \, .
\stp
The conductivity in this expression is defined from the transport integrals
\st
\label{eq:kineticints}
  T\sigma \Delta^{\mu\nu} \equiv  \Delta^{\mu}_{\ph \alpha} \Delta^{\nu}_{\ph \beta}  I^{\alpha\beta}\,,   \qquad  I^{\alpha\beta} \equiv  \int \frac{d^3p}{(2\pi)^3p^0 } \,  \clp^{-1} f_0 \,  p^\alpha p^\beta \, .
 \stp

In the density frame one
proceeds similarly, but uses the lowest order equations in the lab frame 
\st
 \partial_t \muh = -v^{i} \partial_i \muh \, , 
\stp
which yields the form of the first viscous correction 
\st
 \delta f = - \clp^{-1} f_0 (p^i  - p^0 v^i) \, \partial_i \muh \, .
\stp
The diffusive current $J_D^i$ in the density frame is the 
difference between the current and the ideal advection $J^0 v^i$
\st
  J_D^{i} = \int\frac{d^3p}{(2\pi)^3 p^0}   \left(p^i - p^0 v^i\right)  f \, .
\stp
Substituting the 
approximate distribution function $f_0 + \delta f$ 
leads to an appealing positive definite symmetric matrix which determines the mean current\footnote{As discussed below, the matrix $\mathcal K^{ij}$ also determines functional form of the noise in the density frame. In Ref.~\cite{Armas:2020mpr} the matrix $\mathcal K$ is written as $D_{jj}^{\rho\sigma}$. This section shows how this form arises in a microscopic theory.}
\st
\label{eq:Kij1}
  J_D^{i} = \mathcal K^{ij} (-\partial_j \muh)\,,
      \qquad  \mathcal K^{ij} \equiv  \int\frac{d^3p}{(2\pi)^3 p^0}  \, \clp^{-1} f_0 \, (p^i - p^0 v^i) (p^j - p^0 v^j) \,.
\stp
Noting that 
\st
p^i - p^0 v^i =  \left( \Delta^{i}_{\ph\alpha} - v^i \Delta^{0}_{\ph \alpha}  \right)  \Delta^\alpha_{\ph \beta }\, p^{\beta} \,,
\stp
we find that $\mathcal K^{ij}$ has the expected density frame form 
\begin{subequations}
\label{eq:Kij2}
\begin{align}
  \mathcal K^{ij} =& \left(\Delta^{i}_{\ph \alpha} - v^i \Delta^{0}_{\ph \alpha} \right) \left( \Delta^{j}_{\ph \beta} - v^j \Delta^{0}_{\ph \beta }  \right) \,  T\sigma \Delta^{\alpha\beta}  \, , \\ 
   =&  T\sigma \left(\delta^{ij} - v^i v^j\right) \, , 
   \end{align}
   \end{subequations}
where we used the integrals defined in \Eq{eq:kineticints}.  
To summarize this section, we have shown how the form of the dissipative current in the density frame arises in covariant kinetic theory (eqs.~\eqref{eq:Kij1} and \eqref{eq:Kij2}).
This form is not covariant although the underlying kinetic theory is covariant. This arises because
we are trying to approximate the full results of kinetic theory in a specific frame.


\section{Comparison with a kinetic model}
\label{sec:comparison_kinetics}



\subsection{A random walk of massless particles: static case}
In this and the next subsections we will study an analytically tractable  covariant kinetic model in $1+1$ dimensions and investigate how the current in this model approaches the density frame constitutive relation. 
The model consists of massless “particles” moving with the speed of light along a one-dimensional line. The particles experience Poissonian random kicks which changes the direction of their velocities. The dynamical evolution of this system can be mapped to the famous telegraph equation from which an all-orders gradient expansion can be derived. We will then compare the results of this all-order gradient expansion with the predictions from the density frame formalism. 

In this subsection we will analyze this model in a static background case and then in the next subsection consider  a fluid background moving with constant velocity. The results in this subsection are not new and can be found in many places, but they will serve to define the terms for the boosted case.

\begin{figure}[h]
\centering
\includegraphics[width=0.4\textwidth]{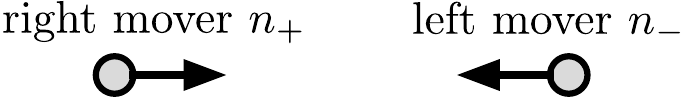}
\caption{Particles moving left and right with density $n_-$ and $n_+$, respectively.}
\label{fig:movers}
\end{figure}

Let us denote the respective densities of the left and right moving particles as $n_-$ and $n_+$ (see Fig. \ref{fig:movers}).

 Due to the random kicks, the particles change direction with the rate  $\Gamma \equiv 1/2 \tau_R$ where $\tau_R$ is the relaxation time. The density of the left/right movers obey the following kinetic equation
\ba
\partial_t n_+ + c\, \partial_x n_+ &=& - \Gamma \left(n_+ - n_-\right)\,,\label{eq:plus}\\
\partial_t n_- - c\, \partial_x n_- &=& - \Gamma \left(n_- - n_+\right)\,.
\label{eq:minus}
\ea
Adding eqs.~\eqref{eq:plus} and \eqref{eq:minus} and rearranging leads to the conservation equation \begin{eqnarray}
    \partial_t n + \partial_x j = 0\,,
\end{eqnarray}
where  $n= n_+ + n_-$ is the density and $j = c(n_+ - n_-)$ is the current. Similarly, subtracting eqs.~\eqref{eq:plus} and \eqref{eq:minus} and rearranging gives the relaxation equation for the current 
\begin{equation}
    \partial_t j + c^2 \partial_x n = - \frac{j}{\tau_R}\,.
\end{equation}
  As usual, this set of first order equations can be expressed as a second order equation for $n$ 
\begin{eqnarray}
\partial_t^2 n+\frac{1}{\tau_R}\partial_t n- c^2 \partial^2_x n=0 \,,
\label{eq:telegraph}
\end{eqnarray}
which is known as the Telegraph equation. 
The exact Green functions associated with this system are known analytically and can be expressed in terms of modified Bessel functions which are given in the Appendix \ref{app:green_functions} for convenience.  For simplicity, we will set $c=1$  for the remainder of this section.

Consider some initial state at $t=0$ (not necessarily in equilibrium) that is specified by two independent functions $n(t=0,x)=n_0(x)$ and $j(t=0,x)=j_0(x)$. Let us further assume that the initial current can be expressed as a gradient expansion of the initial density, with some coefficients $j_0=-\sum_{\ell=0}^\infty \tau_R^{2\ell+1} b_{2\ell+1} \partial_x^{2\ell+1} n_0(x)$. Note that the coefficients $b_{2\ell+1}$ partially characterize the initial state and they are dimensionless by definition. 

By using the exact Green functions, it can be shown that at late times $t\gtrsim \tau_R$ the system obeys a universal dispersion relation
\begin{eqnarray}
j=-\sum_{\ell=0}^\infty c_\ell  \tau_R^{2\ell+1} \partial_x^{(2\ell+1)} n \,,
\label{eq:grad_exp}
\end{eqnarray}
where 
\begin{equation}
  \label{eq:expdecay}
c_\ell=(-1)^{\ell} C(\ell)+ \sum_{m=1}^{\ell+1} e^{-m (t/\tau_R)}P_{\ell,m}(t) \,,
\end{equation}
with $C(\ell)=(2\ell)!/(\ell!(\ell+1)!)$ being the $\ell^{\rm th}$ Catalan number and $P_{\ell,m}(t)$ is a polynomial in $t$ of order $\ell-m+1$ whose coefficients depend on $b_{i\leq \ell}$. It is clear from this expansion that the dependence on the initial conditions decays exponentially for $t \gtrsim \tau_R$, meaning that the system thermalizes and obeys a universal constitutive relation.  The constitutive relation can be expressed as an all-order gradient expansion whose coefficients are given by $(-1)^\ell C(\ell)$. The Catalan numbers are nothing but the Taylor coefficients of the dispersion curve $\omega(k)$
of the diffusion mode:
\begin{equation}
  \label{eq:detequation}
\det \left(\begin{matrix}-i \omega & i k \\ i k &-i\omega+1/\tau_R\end{matrix}\right)=0\Rightarrow \omega(k)=-\frac{i}{2\tau_R}\left(1-\sqrt{1-4k^2\tau_R^2}\right),
\end{equation}
associated with the linear system expanded around $k=0$. 
The dispersion relation has a branch singularity at $k_*=\pm1/(2\tau_R)$.
Furthermore the \emph{large $k$} expansion is consistent with the stability and causality conditions given in Refs. \cite{Heller:2022ejw,GAVASSINO2023137854,Gavassino:2023mad,Hoult:2023clg} where the Telegraph equation was previously analyzed in these terms. 

The additional eigenfrequency from \eq{eq:detequation} is gapped and approaches $\omega(k) = -i/\tau_R + \mathcal O((\tau_R k)^2)$ for $k\rightarrow 0$. The gapped modes are responsible for the exponential decay in \Eq{eq:expdecay}. As we will see in the next section, when the fluid is moving with velocity $v$, the density frame constitutive relation is approached on a short timescale of order $\tau_R/\gamma$, which again reflects the dynamics of gapped non-hydrodynamic modes.

\subsection{A random walk of massless particles: moving fluid}
\begin{figure}[h]
\centering
\includegraphics[width=0.37\textwidth]{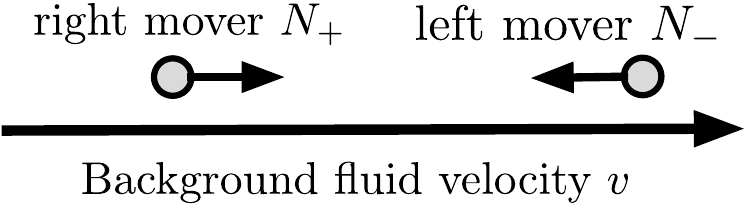}
\caption{Particles moving left and right with density $N_-$ and $N_+$, respectively, in a background fluid moving with velocity $v$.}
\label{fig:boostedmovers}
\end{figure}

Let us now consider the same kinetic model of the previous section in the background of a  fluid moving uniformly with velocity $v$ in the lab frame, as shown in fig.~\ref{fig:boostedmovers}. 

In the lab frame, the density of left and right movers obey
\ba
\partial_t N_+ + c \,\partial_x N_+ &=& - \Gamma_+ N_+ + \Gamma_- N_- \,, \label{eq:boostedplus}\\
\partial_t N_- - c\, \partial_x N_- &=& - \Gamma_- N_- + \Gamma_+ N_+\,, \label{eq:boostedminus}
\ea
where transition rates are 
\ba
\Gamma_+ = \frac{1}{2 \tau_R} \sqrt{\frac{1-v/c}{1+v/c}}\,, \qquad \Gamma_- = \frac{1}{2 \tau_R} \sqrt{\frac{1+v/c}{1-v/c}}\,.
\ea
The appearance of the kinematical factors $\kappa_\pm=\sqrt{(1\pm v/c)/(1\mp v/c)}$ can be understood in the following way:
the right movers in the local rest frame, denoted by the subscript $R$, follow the trajectory $x_R=c t_R$. A Lorentz boost to the lab frame coordinates, $(t,x)$,  leads to the time dilation factor $t=\gamma(1+v/c)t_R=\kappa_+ t_R$ where $\gamma=(1-v^2/c^2)^{-1/2}$ is the Lorentz factor. Since the mean free path time in the local rest frame is $\Gamma^{-1}=2\tau_R
$, the mean free path time in the lab frame is therefore $\Gamma^{-1}_+=\kappa_+ 2\tau_R$. 

Similarly to the static case, by adding and rearranging equations \eqref{eq:boostedplus} and \eqref{eq:boostedminus}, we get
\ba    \partial_t N +  \partial_x J &=& 0 \,,\\ 
  \partial_t J + c^2 \partial_x N &=& \frac{\gamma}{\tau_R} \left(v\,N - J\right)\,,
\ea
where  $N = N_+ + N_-$ and $J = c\,(N_+ - N_-)$ denote the density and the total current respectively. 
Following the density frame approach, we write the current as a sum of the advective and diffusive parts,
\begin{equation}
J=vN+J_D\,.
\end{equation} 
The diffusive part satisfies the relaxation equation 
\begin{equation}
\partial_t  J_D-v\partial_x J_D+\frac{c^2}{\gamma^{2}} \partial_x N+\frac{\gamma}{\tau_R}J_D=0\,.
\end{equation}
For clarity we will set $c=1$ for the rest of this section. 

Let us now consider an initial state given at $t=0$ in the lab frame. Based on our findings in the static case, we expect the system to lose information about the initial conditions for $t \gtrsim \tau_R/\gamma$ and obey a universal gradient expansion
\begin{equation}
J_D=-\frac{1}{\gamma^2}\sum_{n=1}^\infty c_n  \left(\frac{\tau_R}{\gamma}\right)^n \partial_x^{(n)} N \,.
\label{eq:grad_expansion_moving}
\end{equation}
Note that due to the nonzero velocity that explicitly breaks parity, we expect both even and odd terms in the derivative expansion as opposed to the static case, which only contains odd terms. We factored out an overall factor of $\gamma^{-2}$ and the characteristic time scale $\tau_R/\gamma$ explicitly to simplify the expressions for $c_n$. As in the static case, 
the coefficients of the gradient expansion, $c_n$, can be calculated from the dispersion relation 
\ba
\det\left(\begin{matrix}
  -i\omega + v i k &   i k \\
i \gamma^{-2} k & -i\omega-  v i k + \gamma/\tau_R 
\end{matrix}\right)= 0 \,,\\
\omega(k)=\frac{-i}{2(\tau_R/\gamma)}\left(1-\sqrt{1-4k^2(\tau_R/\gamma)^2 - 4 v i k (\tau_R/\gamma)}\right)\,.
\label{eq:dispersion_moving}
\ea
The branch singularity in this case is in the complex plane, $k_*=(\pm 1- i v\gamma)/(2\tau_R)$. Note that the radius of convergence, $|k_*|=\gamma/(2\tau_R)$, grows with $\gamma$ so that the hydrodynamic description applies to modes of wavenumbers $k \lesssim \gamma/\ell_{\rm mfp}$ in the lab frame. Here and below we  define the rest-frame mean-free-path: 
\st
\label{eq:defmfp}
\ell_{\rm mfp}  \equiv  c/\Gamma \equiv 2\tau_R \, .
\stp
The other branch in the dispersion relation is gapped,   $\omega(k) = -i/(\tau_R/\gamma)$  for $k\rightarrow 0$. This branch describes a non-hydrodynamic mode decaying exponentially on a timescale of $\tau_R/\gamma$. Notably, the decay time is \text{not} time-dilated,  but instead time-contracted by a factor of $\gamma$ relative to the static case.  

 The $c_n$ follow from Taylor expanding $\omega(k)$ in \Eq{eq:dispersion_moving} around $k=0$. The even and odd terms are found as
\ba
c_{2n+1}&=&\frac{1}{\gamma^{2n}}\sum_{j=0}^n\frac{(-1)^{n+j}(2v\gamma)^{2j}(2n)!}{(n-j+1)(2j)!\left((n-j)!\right)^2} \,,\nonumber \\
&=&\frac{1}{\gamma^{2n}}\frac{(-4)^n\Gamma(2n+3/2)}{\Gamma(n+3/2)\Gamma(n+2)}\,_2F_1 \left(-1-n,-n,-2n-1/2,\gamma^2\right)\,,
\\
c_{2n+2} &=& 2v \frac{1}{\gamma^{2n}}\sum_{j=0}^n\frac{(-1)^{n+j}(2v\gamma)^{2j}(2n+1)!}{(n-j+1)(2j+1)!\left((n-j)!\right)^2}\,,\nonumber\\
&=& 2v\frac{1}{\gamma^{2n}}\frac{(-4)^n\Gamma(2n+5/2)}{\Gamma(n+5/2)\Gamma(n+2)}\,_2F_1 \left(-1-n,-n,-2n-3/2,\gamma^2\right)\,.
\label{eq:cns_moving}
\ea 
For reference we write down the first few terms below:
\ba
c_1=1, c_2=2v, c_3=- \left(1-5v^2 \right), c_4=-2v \left(3-7v^2 \right), c_5 = 2 \left(1-14v^2+21v^4 \right) \dots
%
\ea
Just like the static case, the large $k$ expansion of the dispersion relation given in  \Eq{eq:dispersion_moving} necessarily satisfy the causality and stability conditions of Refs.~\cite{Heller:2022ejw,GAVASSINO2023137854,Gavassino:2023mad,Hoult:2023clg}.

Putting everything together, we find the gradient expansion in the lab frame as
\ba
J_D&=&-\frac{\tau_R}{\gamma^3}\partial_x N+(1-5v^2)\frac{\tau_R^3}{\gamma^5}\partial_x^3 N-2(1-14v^2+21v^4)\frac{\tau_R^5}{\gamma^7}\partial_x^5 N+\dots
\nonumber\\
&&+v\left[-2\frac{\tau_R^2}{\gamma^4}\partial_x ^2 N+2(3-7v^2)\frac{\tau_R^4}{\gamma^6}\partial_x ^4 N+\dots\right]\,.
\label{eq:grad_expansion_moving_J}
\ea
Let us discuss this result.
The leading term in \Eq{eq:grad_expansion_moving_J} agrees with the prediction from the advection-diffusion equation in the density frame. In some sense, the higher order corrections encode the underlying microscopic dynamics and they restore Lorentz causality as can be seen from the large $k$ expansion, for instance. Of course Lorentz causality is violated if the gradient expansion is truncated at any finite order. 

\begin{figure}
    \centering
    \includegraphics[width=0.495\linewidth]{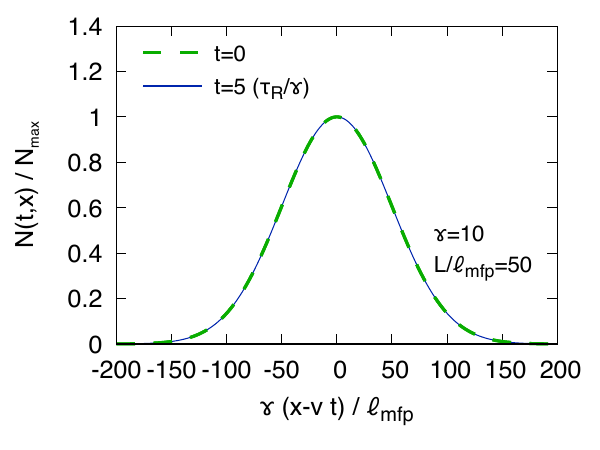}
    \includegraphics[width=0.495\linewidth]{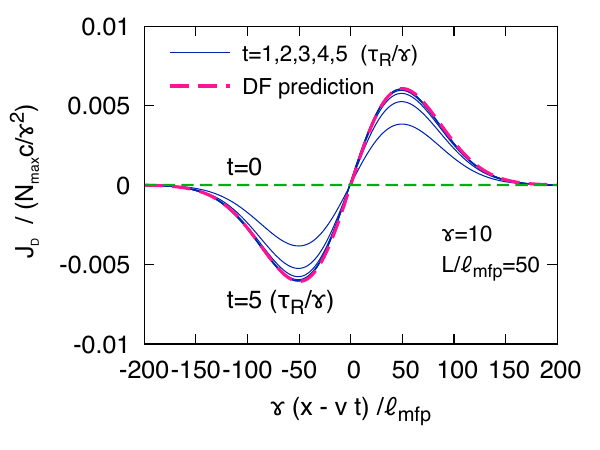}
    \caption{(a) The evolution of a Gaussian drop of charge in the lab frame in a kinetic model of \Sect{sec:comparison_kinetics} for a moving fluid with Lorentz factor $\gamma=10$. The Gaussian has a lab frame width of $\sigma=L/\gamma$ where $L$ is fifty times the rest frame mean-free-path, $L=50\,\ell_{\rm mfp}$. We are studying  a very short time  interval,  $\Delta t = 5\,\tau_R/\gamma$, i.e. a time interval of order the rest frame relaxation-time \emph{divided} by $\gamma$. Over this interval the drop is advected,  but the diffusion of the drop is negligible. (b) The relaxation of the lab frame diffusive current in the kinetic model over the same short time interval as (a) starting from $J_D=0$ at $t=0$. The current at different times is compared to the leading order density frame prediction, $J_D = -(D/\gamma^3) \partial_x N$, which is essentially time independent.
    \label{fig:thermalization} }
\end{figure}

One might ask when the physics of the covariant kinetic model is captured by  the density frame diffusion framework. We first require that the system reaches local thermal
equilibrium and therefore can be described by the universal constitutive
relation. This happens in a timescale of order $t\sim \tau_R/\gamma$, which is set by the inverse frequency of the gapped modes. This is
illustrated in Fig. \ref{fig:thermalization} where at $t=0$ we initialize a
Gaussian drop of charge, $N_0(x)=N_{\rm max} \exp(-x^2/2\sigma^2)$, and set the 
initial diffusive current to zero in the kinetic model, $J_D=0$.
For the test case shown we 
took a fluid with $\gamma=10$ and set  $\sigma=L/\gamma$ with  $L/\ell_{\rm mfp}=50$.   
The left figure  shows the time evolution of the charge density in 
the lab frame.  In the short period of time considered in the figure $\Delta t\sim \tau_R/\gamma$, the charge does not have enough time to diffuse and it is simply advected  by the background flow. 
The right figure shows the evolution of
the lab frame diffusive current  over the same time period.   Clearly the diffusive 
current in the kinetic model relaxes on a time of order $\tau_R/\gamma$ to a steady state, and at $t=5\tau_R/\gamma$ the steady state agrees with the leading derivative term of density frame gradient expansion given in \Eq{eq:grad_expansion_moving_J} to a very good precision. 

The relaxation timescale $\tau_R/\gamma$ is easily understood.
Indeed, the mean collision times of right and left movers are of the order $\gamma \tau_R$ and $\tau_R/\gamma$ respectively.
In equilibrium,  where $J=Nv$, the number of right and left movers is 
\st
    N_+^{\rm eq} = N \frac{1 + v }{2}\, ,    \qquad N_{-}^{\rm eq} = N \frac {1-v}{2} \, , 
\stp
and thus for $v\rightarrow 1$  there are almost $N$ right movers and of order $\sim N/\gamma^2$ left movers in the equilibrium sample.
If all the particles are initially left movers,  then in a time of order $\tau_R/\gamma$, these initial particles will scatter with probability one, reaching approximate equilibrium with  $N_+\simeq N$. Similarly, if all the particles are initially right movers,  then in a time of order $\tau_R/\gamma$,   $N/\gamma^2$ of these initial particles will scatter,  generating a yield of left movers commensurate with equilibrium, $N_- \sim N/\gamma^2$. Thus, the typical equilibration time is of order $\tau_R/\gamma$.

\begin{figure}
    \centering
    \includegraphics[width=0.495\linewidth]{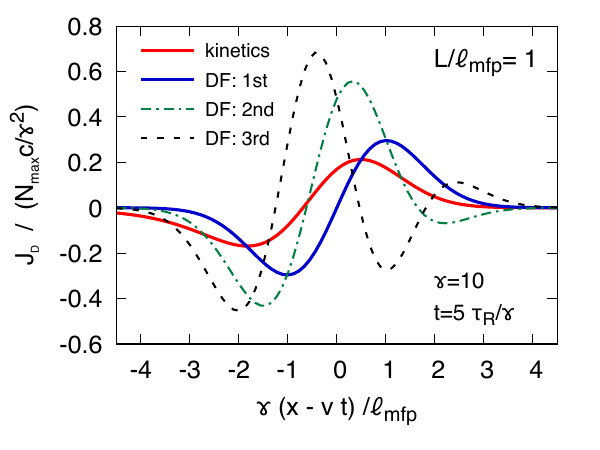}
    \hfill
    \includegraphics[width=0.495\linewidth]{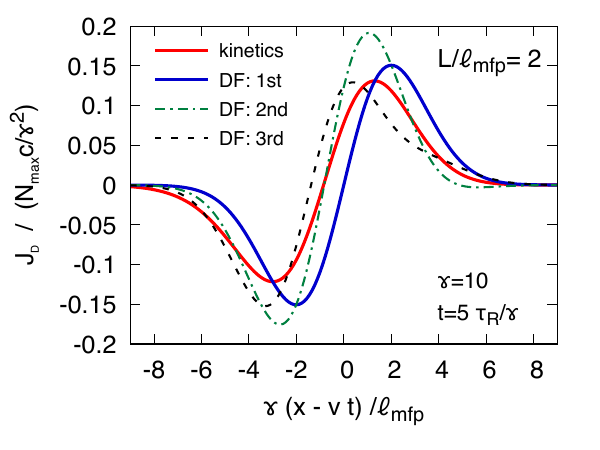}
    \includegraphics[width=0.495\linewidth]{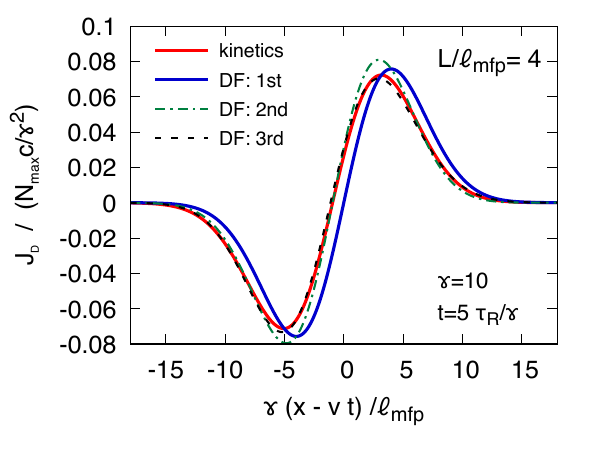}
    \hfill
    \includegraphics[width=0.495\linewidth]{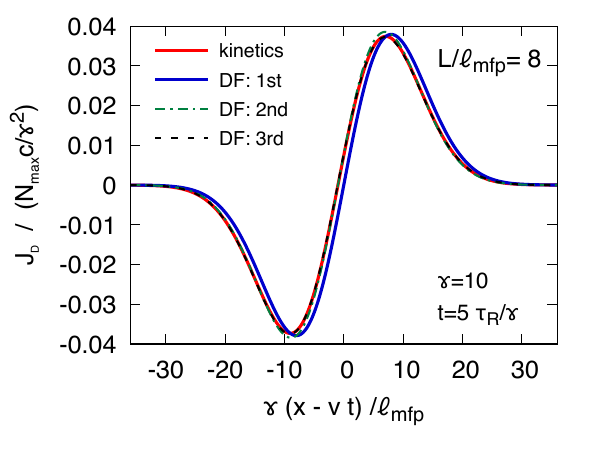}
    \caption{Comparison of the first three orders of density frame gradient expansion (DF: 1, 2, 3 given in \Eq{eq:grad_expansion_moving_J}) with the kinetic model, for the same setup as \Fig{fig:thermalization} but for different rest frame system sizes $L$.  For $L=1\, \ell_{\rm mfp}$ the gradient expansion does not converge at all, however, the gradient expansion becomes increasingly accurate with the increase in system size.}
    \label{fig:convergence}
\end{figure}

Of course, the applicability of the gradient expansion depends on the size of
the system as well, and we expect it to break down for smaller systems. In Fig.
\ref{fig:convergence} we show the diffusive current for the same setup as above with
$\gamma=10$ at $t=5\tau_R/\gamma$, but with different initial Gaussian
sizes ranging from $L= (1\ldots 8)\, \ell_{\rm mfp}$ in the rest frame. From the figure we see that for a small system when $L$  is one mean-free-path, the gradient expansion does not converge at all. As
the system size gets larger, the gradient expansion becomes more and more
accurate.
Note that the leading term in the gradient remains  qualitatively well behaved even for very small systems.
Remarkably, the convergence starts already when the system size is only $3$ or $4$ mean-free-paths long, where the second and third order terms in the gradient expansion capture the physics quite accurately. Notably in this intermediate region, the effect of the second order term is clearly seen  by comparing the red (exact) and blue (first order) curves in the lower left figure. This skewness is due to the background motion, and is absent in the static case where parity is unbroken.

The convergence of the gradient expansion can be analyzed more rigorously. Let us consider the time interval $\tau_R/\gamma \ll t \ll \tau_R \gamma^3$ where the system is in local thermal equilibrium but the initial wave packet has not diffused yet. In this regime $N(t,x)\approx N_0(x)$, therefore, the gradient expansion,  \Eq{eq:grad_expansion_moving}, becomes
\begin{equation}
J_D\sim -\gamma^{-2}\sum_{n=1}^\infty c_n \left(\frac{\tau_R}{\gamma}\right)^{n} \partial_x^{(n)} N_0 \sim -\gamma^{-2}\sum_{n=1}^\infty c_n \left(\frac{\tau_R}{\sqrt{2}\gamma \sigma} \right)^n e^{-\frac{x^2}{2\sigma^2}} H_n(x) \,,
\end{equation}
where $H_n(x)$ denotes the $n^{th}$ Hermite polynomial that follows from the gradient expansion of the Gaussian wave-packet with the width $\sigma$. We do not keep track of the overall magnitude of the initial density as it is inconsequential for our argument.  It is straightforward to observe that the coefficients, $c_n$, grow exponentially in $n$ \footnote{This can be seen from the dispersion relation,  \Eq{eq:dispersion_moving}, whose Taylor coefficients around $k=0$ are $c_n$.}. Given that asymptotically, the Hermite polynomials grow factorially as $H_n\sim\Gamma(n/2)$ (apart from the finite number of points where they vanish), the gradient expansion is an asymptotic expansion. The effective coupling constant of this asymptotic series is $\tau_R/(\sigma\gamma)$. A well known property of an asymptotic series that grow as $g^n \Gamma(n/2)$ is that it starts to diverge at order $n^*\sim 2/g^2$. According to the optimal truncation procedure a la Poincar\'e, the gradient expansion can be directly summed up to $n^*$ terms before the series start to diverge. 
This result implies that when the system size is comparable to mean-free-path, namely $\sigma = c \tau_R/\gamma$ for some constant $c$, we get $n^*=1$ and the optimal truncation breaks down.

Furthermore in the ultra-relativistic limit, $v\rightarrow1$, the gradient expansion coefficients significantly simplify; $c_n=2^{n-1}+{\cal O}(\gamma^{-1})$ and we find the asymptotic behavior of the gradient expansion to be
\begin{equation}
J_D \sim -\sum_n (-1)^n \left(\frac{2\sqrt{2} \tau_R}{\gamma\sigma}\right)^n \Gamma\left(\frac{n+1}{2}\right)e^{-\frac{(x-t)^2}{2\sigma}}\cos(\sqrt{2n}(x-t)-n\pi/2)\,.
\end{equation}
Therefore the breakdown of the optimal truncation occurs when $\sigma=2\tau_R/\gamma$, confirming our heuristic argument earlier.

\section{Stochastic dynamics}
\label{sec:stochastic_dynamics}
\subsection{Noise in the density frame}

In this section we will add noise to the density frame diffusion equation and study the stochastic dynamics of a boosted fluid.
The dissipative current in the density frame take the form
\st
J_{D}^{i} = -T\sigma^{ij} \partial_j \muh + \xi^i \,,
\stp
where $\xi^i$ is the noise.
For the stochastic process to equilibrate 
to the probability distribution determined by the entropy  of the system,
\st
P[N] \propto \exp\left(\mathcal S[N]  \right)  \propto \exp\left( - \int \dd^3x \, \frac{\beta^0 }{2 \chi^{00}} N^2  \right)\,,
\stp
the noise must respect the fluctuation-dissipation theorem
\st
   \llangle \xi^i(t,\x) \,\xi^j(t', \x') \rrangle  = 2 T \sigma^{ij} \,   \delta^3(\x - \x') \, \delta(t-t') \, . 
\stp
In \app{app:spacetimedepend} we describe how the noises added to the system  should be generalized when the fluid velocity depends on space and time.

The form of the noise matrix in the density frame can also be found by algebraically manipulating the current in the Landau frame.  In the Landau frame, we have
\st
  J^{\mu} = n_{\LF} u^{\mu} + j_\DLF^{\mu} + \xi^{\mu}_\LF \,,
\stp
where the noise is orthogonal to $u^{\mu}$  and satisfies
\st
   \llangle \xi^{\mu}_\LF \, \xi^{\nu}_\LF \rrangle = 2 T\sigma \, \Delta^{\mu\nu} \delta^4 (x -y) \,.
   \label{eq:landau_noise_covariance}
\stp
Rearranging the Landau frame variables into the density frame form, we find
\st
  J^{\mu} = (N, N v^i + J_D^i + \xi^i)\, ,
\stp
where the density frame noise is 
\st
 \xi^{i} = \xi_\LF^i - v^i \xi_\LF^0 = (\Delta^i_{\ph \alpha} - v^i \Delta^0_{\ph \alpha})\, \xi^{\alpha}_\LF \,.
\stp
Computing the covariance  of the density frame noise  using \Eq{eq:landau_noise_covariance} yields
\begin{align}
  \llangle \xi^i(x) \xi^j(y) \rrangle =&
 (\Delta^i_{\ph \alpha} - v^i \Delta^0_{\ph \alpha}) 
 (\Delta^j_{\ph \beta} - v^j \Delta^0_{\ph \beta})\,  2 T \sigma \Delta^{\alpha \beta} \delta^4(x -y) \,, \\
  =&  2 T\sigma^{ij}  \delta^4(x -y)\,.
\end{align}
Thus the form of the noise in the density frame can be straightforwardly found from the Landau frame definitions. 
%
\subsection{The Metropolis algorithm for stochastic equations}

Next we discuss how the dissipative stochastic dynamics can be simulated  using a Metropolis algorithm, rather than directly discretizing the Langevin dynamics. As discussed in the introduction, the approach has the advantage that detailed balance is maintained irrespective of the time step $\Delta t$.

To understand the method, consider the one-dimensional Brownian motion of a particle in a potential $U(q)$. The  Brownian particle evolves  in phase-space as
\begin{subequations}
     \label{eq:BrownianMotion}
\begin{align}
\frac{dq}{dt}  + \left\{\mathcal H , q \right\}  =& 0  \, ,  \\
\frac{dp}{dt}  + \left\{\mathcal H, p \right\}  =& - \eta \left(\frac{\partial \mathcal H}{\partial p}\right)  + \xi \,,  \qquad \qquad  \llangle \xi(t) \xi(t') \rrangle = 2 T\eta \, \delta(t- t')\,,
     \end{align}
     \end{subequations}
 where the free energy of the  particle with  momentum $p$  and position $q$ is 
\st
    \mathcal H(q,p) = \frac{p^2}{2m } + U(q) \,.
\stp
Here $\eta$ is the drag coefficient and  the drag force is proportional to the velocity, $\partial \mathcal H/\partial p = v$, i.e. the variable thermodynamically conjugate to $p$.
The noise is chosen so that the system evolves to the equilibrium probability distribution
$P(q, p)  \propto e^{-\beta\, \mathcal H(q,p)}$.  

A natural way to simulate the dynamics is to use operator splitting,  first  setting the right hand side of \Eq{eq:BrownianMotion} to zero and taking a symplectic step. Ideally, this step should be done with a symplectic integrator which preserves the phase-space volume.  The symplectic update is followed by a dissipative step such as the Metropolis update discussed below, which respects detailed balance.  Together, the two steps correctly evolve \Eq{eq:BrownianMotion} over a time  $\Delta t$.

In the  Metropolis update algorithm  over a time interval $\Delta t$,
one makes a proposal 
\st
\label{eq:Brownnoise}
        p \rightarrow p + \Delta p    \,,  \qquad \qquad \Delta p = \sqrt{2 T\eta \Delta t} \, \mathfrak{e} \,,
\stp
where $\mathfrak{e}$ is a random number of variance one. 
Then the change in free energy in the proposed step is 
\st
  \beta \Delta \mathcal H = \beta \left(\mathcal H(p + \Delta p) - \mathcal H(p) \right) \simeq  \beta \frac{\partial \mathcal H}{ \partial p  } \, \Delta p \,.
\stp
In a Metropolis approach if $\Delta \mathcal H$ is negative then the proposal is accepted; if $\Delta \mathcal H$ is positive then the proposal is accepted with probability $e^{-\beta \Delta \mathcal H} \simeq 1 -  \beta \Delta \mathcal H$.
Because of the asymmetry between gain and loss rates, the particle will experience drag in addition to the noise added in \Eq{eq:Brownnoise}. 
It is straightforward to see that the mean  momentum transfer $\Delta p$  from the  Metropolis step  is 
\st
   \llangle \Delta p \rrangle \simeq  \int_{-\infty}^{\infty} \dd \mathfrak{e} P(\mathfrak{e}) \left[\Theta(-\Delta \mathcal H)  + \Theta(\Delta \mathcal H) \left(1 - \beta \Delta \mathcal H \right) \right]  \Delta p  \simeq   - \eta \frac{\partial H}{\partial p} \Delta t \,,
\stp
reproducing the mean drag in the Langevin equations of motion. A similar computation shows that $\llangle (\Delta p)^2 \rrangle = 2 T \eta \, \Delta t$, indicating that the Metropolis algorithm correctly reproduces the drag and noise of the Langevin evolution.

%

\subsection{The advection diffusion equation from the Metropolis algorithm }
\label{sec:num_implement}
Now we will show how the same Metropolis algorithm can be used to simulate
the Langevin updates for the relativistic advection-diffusion equation in the density frame.
 The continuum equation we would like to solve is 
\st
\partial_t N + \partial_i (N v^i)  +  \partial_i (J_D^i + \xi^i) = 0\,,
\label{eq:continuum_equation}
 \stp
 where the dissipative part is
 \st
   J_D^{i} = -T\sigma \left( \delta^{ij} - v^i v^j\right) \partial_j \muh\,.
 \stp
 For simplicity we will limit the discussion  to two spatial dimensions. We also 
 have kept the fluid velocity constant, discussing the more general case in \app{app:spacetimedepend}.

 As in the Brownian motion example an operator splitting approach is adopted, we first  solve the advection equation 
 \st
     \partial_t N + \partial_i (Nv^i) = 0 \,,
 \stp
 which captures the symplectic dynamics in this case.   For the advective step we 
 adopt the Kurganov-Tadmor (KT) central scheme using a second order spatial discretizati on~\cite{Kurganov:2000ovy,Zanna:2002qr,Schenke:2010nt}.  Given the stochastic nature of the simulation, we turned off limiters such as min-mod or WENO based limiters. This choice and possible alternatives should be reexamined in the future. For the time integration we use a second order Total Variation Diminishing (TVD)  Runge Kutta method implemented in the {\tt PETSc} library~\cite{Gottlieb2009,petsc-web-page}. We use a fixed time step $\Delta t=0.5\, \Delta x/c$  in the numerical experiments presented below. 
 We note that both advection and ideal relativistic hydrodynamics  have a symplectic structure, which can be derived from Poisson brackets between the conserved charges~\cite{DZYALOSHINSKII198067}.  However, the KT scheme with the TVD time discretization is not a symplectic integrator. It would be interesting to explore symplectic integrators for ideal hydrodynamics when physical dissipation (which naturally leads to TVD property)  is incorporated in subsequent steps.

 After taking an advective step, we propose random transfers of charge between the fluid cells with appropriate variances. The charge transfers are accepted or rejected according to the statistical weight, $\exp (\Delta \S)$.  The procedure parallels the Brownian motion example of the previous subsection and reproduces the mean diffusive current as well as the noise.  In the next paragraphs we will explicitly list the algorithm and verify this claim.

The simulation is discretized on a two dimensional lattice with a finite volume discretization and fixed lattice spacing.  The lattice metric is 
\st
  ds^2 = a_x^2  \left(\Delta I_x \right)^2 + a_y^2 \left(\Delta I_y \right)^2\,,
\stp
with $a_x$ and $a_y$ being lattice spacing, and $I = \left(I_x, I_y \right)$ denotes the (integer) lattice coordinates.
The volume of a  fluid  cell is $V_0 = \sqrt{g} = a_x a_y$  and 
the charge in the $I^{\rm th}$ cell is  $\N_I = V_0 N_I$.
The entropy of the system takes the form 
\st
     \mathcal S[\N] = \mathcal S_1  - \sum_{I}  \frac{ {\N}^{\,2}_{I} }{2 T\chi u^0 V_0}  \,,
\stp
and derivatives of $\mathcal S$ with respect to $\N_I$ determine the chemical potential
\st
  \frac{\partial \S}{\partial \N_I} =  -\muh_I \,,
  \label{eq:dSdN}
\stp
where  $\muh_I = \N_I/T\chi u^0 V_0$.

\begin{figure}
   \centering
 \includegraphics[width=0.3\textwidth]{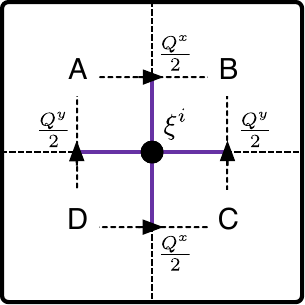}
 \caption{Discretization of Metropolis proposals. The noise $\xi^i$ ``lives" on the corner of the computational cells, $A$, $B$, $C$, $D$. The charge transfers between cells are given in \Eq{eq:qtransfer}.}
 \label{fig:discretization}
   \end{figure}

The layout of the grid is shown in  Fig.~\ref{fig:discretization}. 
We imagine a stochastic current living at the corner of the computational cells  with covariance
\st
\label{eq:noisecorvar}
     \llangle \xi^i \xi^j \rrangle =  \frac{2 T \sigma}{\Delta t V_0} \left(\delta^{ij} - v^i v^j\right)\,,
\stp
where $\Delta t$ is the time step.
One way to produce this noise
 is to generate noises parallel and perpendicular to the fluid velocity, $\xi_\parallel$ and $\xi_\perp$,  with variances
\begin{align}
  \label{eq:proposal_method}
   \xi_\parallel =& \sqrt{ \frac{2 T \sigma}{\Delta t V_0} (1 - v^2) } \;  {\mathfrak e}_\parallel \,, \\
  \xi_\perp =& \sqrt{ \frac{2 T \sigma}{\Delta t V_0}  } \;  {\mathfrak e}_\perp \,,
  \end{align}
where ${\mathfrak e}_\parallel$ and ${\mathfrak e}_\perp$ are random numbers with zero mean and unit variance.  Then a rotation gives a proposal 
with the expected variance in \Eq{eq:noisecorvar}.

The proposed charge transfer in the $x$ and $y$ directions are
\st
  Q^i = \xi^i A_i  \Delta t  = \xi^i V_0 \Delta t/a_i\,.   \qquad \mbox{(no sum)} 
\stp
Here and for the rest of this section no sum is implied by repeated indices.
The variance of the proposed charge transfers is 
\st
   \llangle Q^i Q^j  \rrangle  = \frac{2 T\sigma \Delta t V_0}{a_i a_j} \left( \delta^{ij} - v^i v^j \right)\,,
\stp
leading to the proposed updates for the cells $A$, $B$, $C$, and $D$ (see \Fig{fig:discretization})
\begin{subequations}
  \label{eq:qtransfer}
\begin{align}
   \N_A \rightarrow&  \N_A + \Delta \N_A \equiv  \N_A - \frac{Q^x}{2} + \frac{Q^{y}}{2} \,, \\
   \N_B \rightarrow&  \N_B + \Delta \N_B \equiv \N_B  + \frac{Q^x}{2} + \frac{Q^{y}}{2} \,, \\
   \N_C \rightarrow&  \N_C + \Delta \N_C \equiv \N_C + \frac{Q^x}{2} -  \frac{Q^{y}}{2}  \,, \\
   \N_D \rightarrow&  \N_D + \Delta \N_D \equiv \N_D - \frac{Q^x}{2} -  \frac{Q^{y}}{2}  \,.
\end{align}
\end{subequations}
Then we compute the change in the entropy for a proposal, using \eqref{eq:dSdN}, which reads
\begin{align}
   \Delta \S =& \sum_{U=A,B,C,D} \S[\N_U + \Delta \N_U]  - S[\N_U]\,,  \\
   \simeq&  -\half (\muh_B + \muh_C - \muh_A - \muh_D) \, Q^x
                 - \half (\muh_A + \muh_B - \muh_C - \muh_D) \, Q^y\,.
\end{align}
Formally, one can also write this as 
\st
    \Delta \S \simeq  - \sum_i \partial_i \muh \, a_i Q^i \,,
\stp
where it is understood that, for instance,
$\partial_x \muh  \equiv (\muh_B + \muh_C - \muh_A - \muh_D)/2a_x$.
Then the probability of accepting the proposed update is 
\st
P_{\rm accept}(\xi) = \theta(\Delta \S) + \theta \left(-\Delta \S\right) \left(e^{\Delta \S} - 1\right) \simeq 1 + \theta(-\Delta \S) \, \Delta \S \,.
\stp
Thus, the mean charge transfer in the Metropolis step is given by
\begin{align}
  \llangle Q^i\rrangle_{\rm accept} \simeq& 
   \int \dd^2 \xi\;  P(\xi) \;  Q^i\, \left(1 + \theta \left(-\Delta \S \right) \Delta \S\right)\,,\nn \\
   \simeq &   - \frac{1}{2} \sum_j \partial_j \muh  \, a_j  \llangle Q^i Q^j \rrangle\,, \nn \\
   =& - \frac{T \sigma  V_0 \Delta t }{a_i }  \sum_j (\delta^{ij} - v^i v^j) \partial_j \muh \,.
   \label{eq:mean_charge_transfer}
 \end{align}
The factor of a half in the second line arises because we are only integrating over proposals where $\Delta \S < 0$. 
Dividing by  $A_i \Delta t = V_0 \Delta t/a_i$, we find the mean current from the metropolis step
that is consistent with the expected form of the diffusive current in the density frame
 \begin{align}
    \llangle J^i_D \rrangle \simeq& -T \sigma  \sum_j (\delta^{ij} - v^i v^j) \partial_j \muh \,.
   \label{eq:mean_current}
   \end{align}
To summarize, the full procedure consists of an ideal advective step,  followed by a diffusive step. In the diffusive step we step over the lattice by two's, first updating the group of cells $A$, $B$, $C$, $D$ and then proceeding to the next independent group of four cells. The updates are independent of each other and can be done in any order.  This covers one quarter of the lattice. We then loop through the remaining
three corners in a similar way to complete the diffusive steps. In fact, to eliminate the potential bias, the order of the four corners which are updated is randomly shuffled in each diffusive step. 

There are many choices and questions here which can be studied in future work.  For instance, it is not necessary to take one diffusive step per advective step. In fact, in order to be closer to the Langevin limit we take $400$ diffusive steps per advective step. This is quite a large number and guarantees that the rejection probability $r$ approaches zero.  It is only in the asymptotic limit  $r\rightarrow 0$ that the Metropolis updates are fully equivalent to the Langevin simulations. In our numerical experiments $r=0.003$. It would be helpful to explore the approach to the Langevin limit in greater detail, which allow for better algorithms that capture the interplay between the symplectic and dissipative dynamics.  

\subsection{Equilibrium correlation functions}

As a first test of the stochastic dynamics we will compute the correlation function of charges advecting and diffusing in a two dimensional fluid moving with a fixed velocity, 
${\bf v} = v\, (\cos\theta, \sin\theta)$. In our test case we treated a fluid moving at
$v=0.8\,{\rm c}$ at an angle of $\theta = \pi/6$ and used a lattice of $L^2=128^2 \,a^2$ where 
$a=a_x=a_y$ is the lattice spacing.

\subsubsection{Physical considerations}
Three dimensionful parameters in the simulation can be set to unity, setting
our units of  space,  time,  and energy.  We choose the lattice length and the
speed of light to be one, $a=c=1$.  The variance of the charge in a fluid cell
$\llangle \N^{\,2} \rrangle = T\chi u^0 a^d$  may also be set to unity, where $d=2$
is the number of spatial dimensions. All physical quantities  can be expressed
in terms of $a$, $c$ and $T\chi u^0$.

The ``mean free path'' of the system $\ell_{\rm mfp}$ is defined through the diffusion coefficient 
$D \equiv \tfrac{1}{3} \ell_{\rm mfp} c$.  The mean free path in units of the lattice spacing $\ell_{\rm mfp}/a$ is a dimensionless parameter, which can only be fixed through physical considerations.   We are only interested in modes where $k \ell_{\rm mfp} \ll 1$, as wave-numbers of 
order  $1/\ell_{\rm mfp}$ have been integrated out of the hydrodynamic effective theory.  Thus, we set $a = \ell$  which cuts off the wave numbers in the simulation at a reasonable value.
Hence, long wavelength modes on the lattice
are well described by the continuum description and the diffusion equation, while modes of
order the lattice spacing are neither resolved nor adequately described by the diffusion equation.



In modeling  the charge fluctuations we have ignored the discrete nature of the charge carriers, neglecting shot noise. Consider a field theory with a finite number of fields and assume that charge susceptibility is of order  
$T\chi  \sim  e^2 (T/\hbar c)^d$,  where $e$ is the elementary charge. This is the case for the electric charge susceptibility of QCD at high temperatures.
The variance of the charge within a fluid cell in units of the elementary charge $e$ is 
\st
\frac{a^d}{e^2} \llangle \delta N^2 \rrangle = \frac{a^d}{e^2} T\chi u^0  \sim \left( \frac{a T}{\hbar c} \right)^d  \, .
\stp
If the theory is weakly coupled $\ell_{\rm mfp}  \gg (\hbar c/T)$,
then this variance is large for $a\sim \ell_{\rm mfp}$,   and it is 
appropriate to treat the charge and charge fluctuations using  continuous variables
even for short wavelengths,  $k a \sim 1$.
When simulating weakly coupled fluids,  errors will still arise from the space-time discretization  and from using the diffusion equation for $k\ell_{\rm mfp} \sim 1$, but not arise from treating charge as a continuous variable.
In a strongly coupled field theories  where $\ell_{\rm mfp} \sim \hbar c/T$ and $\chi \ell^d_{\rm mfp}/e^2 \sim 1$, the variance of a fluid cell in units of $e^2$ is of order unity  and the discretized hydrodynamic theory does not capture the quantized charge fluctuations (or shot noise) at the scale of the mean free path. This error is the same order of magnitude as the discretization and modeling errors made in the weakly coupled case.
(In strongly coupled, but large $N_c$ field theories, the susceptibility is of order $T\chi \sim N_c \,e^2 (T/\hbar c)^d$ and  shot noise is always negligible.)
Finally, in theories where the susceptibility is very small $\chi \ell_{\rm mfp}^d/e^2 \ll 1$, it should be possible to include  shot noise systematically into the hydrodynamic description by making discrete Poissonian proposals for the charge transfers  between fluid cells. However,  we have adopted continuous charge transfers here and leave this regime for future work.




\subsubsection{Numerical results}
The density-density correlation function in the simulation is
\st
  C_{NN} (t-t', {\bf k}) \equiv \int \dd^3x \, e^{-i {\bf k}\cdot \x} \llangle N(t, {\bf x}) N(t', {\bf 0} ) \rrangle \equiv \frac{1}{V} \llangle N(t,{\bf k}) N(t', -{\bf k} ) \rrangle  \, . 
  \label{eq:simulated_corre_fn}
\stp
From the density frame equation of motion 
\st
    \partial_t N + \partial_i (N v^i) -  \partial_i \left(D^{ij} \partial_j N \right) + \partial_i \xi^i = 0 \, ,
\stp
the expected correlation function can be computed straightforwardly.
Indeed, one can solve for $N(t, {\bf k})$  in terms of $\xi(t, {\bf k})$   
 \st
    N(t, {\bf k}) = \int^t_{-\infty} dt^\prime\,   (-i k_m \xi^m(t^\prime, {\bf k}) )\, e^{-i{\bf v}\cdot {\bf k}\, (t - t^\prime)  } e^{-D^{ij} k_i k_j \, (t- t^\prime) }  \, . 
 \stp
 Squaring this expression and averaging over the noise determines
 the expected form of the density frame correlation function 
 \st
 \label{eq:correlationfcn}
   C_{NN}(t-t', {\bf k}) =     T\chi u^0 \cos({\bf v} \cdot {\bf k} \, (t -t')) \exp( -D^{ij} k_i k_j \, |t-t'| )\,.
 \stp

\begin{figure}
    \centering
    \includegraphics[width=0.495\linewidth]{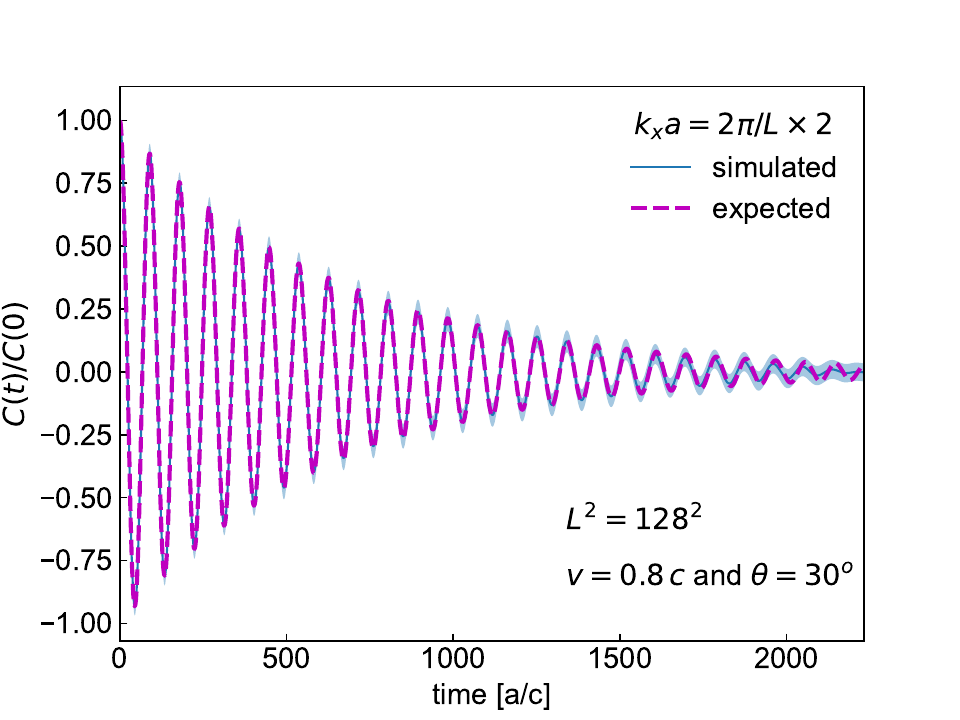}
    \hfill
    \includegraphics[width=0.495\linewidth]{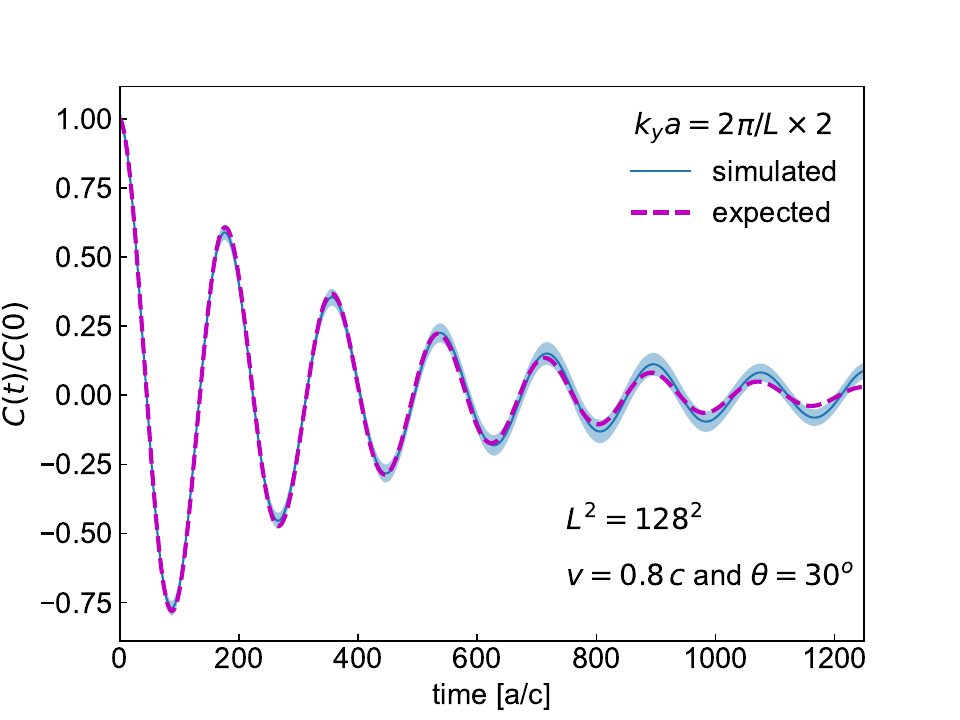}
    \includegraphics[width=0.495\linewidth]{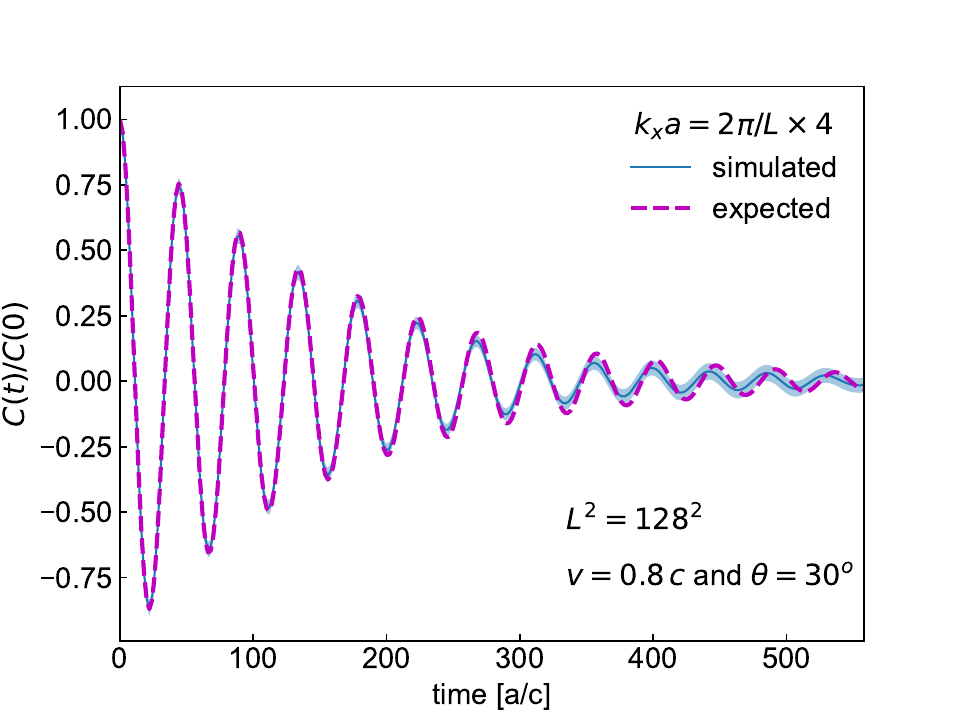}
    \hfill
    \includegraphics[width=0.495\linewidth]{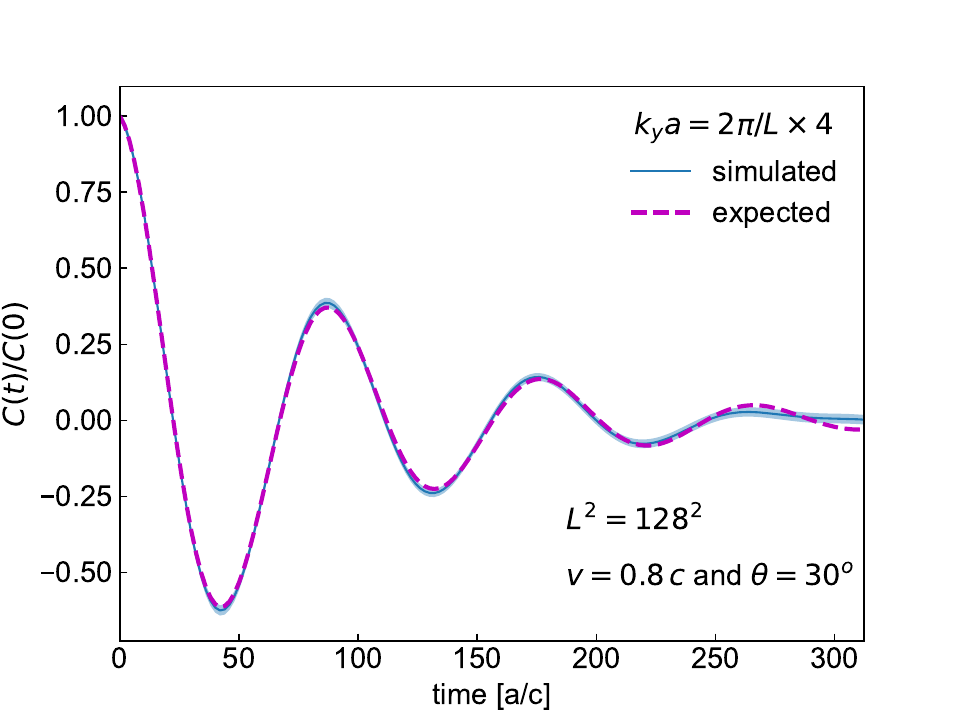}
    \includegraphics[width=0.495\linewidth]{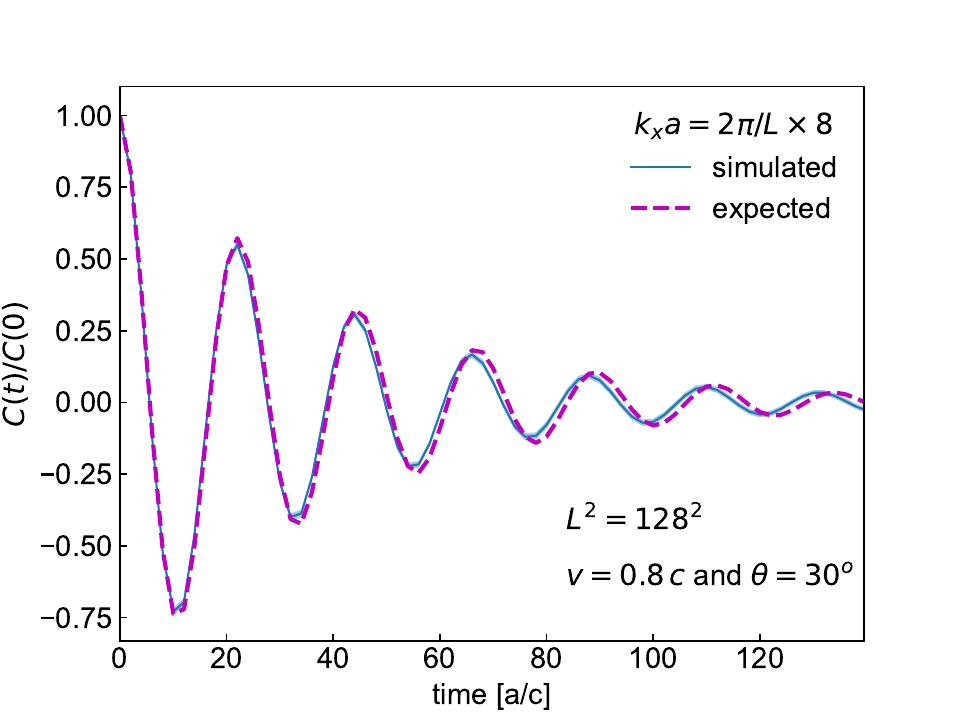}
    \hfill
    \includegraphics[width=0.495\linewidth]{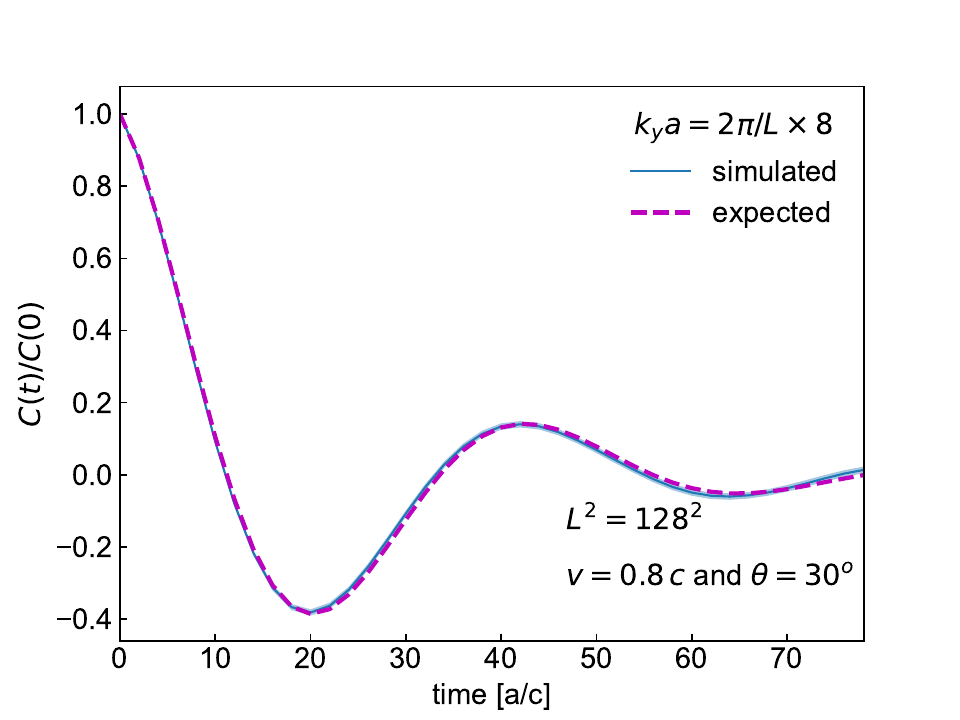}
    \caption{Comparison between the expected \eqref{eq:correlationfcn} and simulated  density-density correlation functions in the density frame as a function of time (in lattice units) for various wave numbers. 
    The fluid is moving with velocity $v=0.8c$ at an angle of $30^\circ$ above  the $x$ axis.  The left  plots have wave numbers $(k_x, k_y) = (k_x, 0)$ for  various values of $k_x$. Similarly, the right plots have $(k_x, k_y) = (0, k_y)$ for various values of $k_y$.}
    \label{fig:correlationfunctions_comparison}
\end{figure}

Figure~\ref{fig:correlationfunctions_comparison} shows a comparison between the expected \eqref{eq:correlationfcn} and simulated correlation functions for different wave numbers in the fluid.  Examining \Eq{eq:correlationfcn} we see that the oscillations reflect the advection of a sinusoidal wave, while the exponential decay is controlled by the diffusion matrix $D^{ij}$. Naturally, the longest wavelengths (smallest wave-numbers) have the slowest decay, and this is clearly seen from the trends in \Fig{fig:correlationfunctions_comparison}. The diffusion matrix takes the form
\st
  D^{ij} = \frac{D}{\gamma^3}
      \vh^i \vh^j + \frac{D}{\gamma} \left(\delta^{ij} - \vh^i \vh^j \right) \,,
\stp
and thus Fourier modes which are parallel to the flow velocity will decay slowly relative to the transverse modes, i.e. at rates $\sim Dk^2/\gamma^3$ and $\sim Dk^2/\gamma$ respectively. The curves with $k_y=0$ and $k_x$ finite (the left hand side of \Fig{fig:correlationfunctions_comparison}) are more aligned with the flow than the curves with $k_x=0$ and $k_y$ finite (the right hand side of \Fig{fig:correlationfunctions_comparison}), and thus the right plots show a stronger exponential decay, confirming this expectation.

\section{Discussion} 
\label{sec:discussion}

We have shown how  the Metropolis algorithm can be  
adapted to simulate stochastic relativistic advection diffusion equation in two dimensions.  In a companion paper we will describe how the framework can be further extended to stochastic  viscous hydrodynamics  in general relativity. 

 The algorithm is simple: \begin{enumerate*}[label=(\roman*)] 
   \item  take an ideal advective step,  \item  make a proposal  to randomly transfer the conserved charges  between computational cells, and finally
   \item accept or reject the proposed transfers based on how they change the entropy of the fluid in a Metropolis-Hastings accept-reject step. \end{enumerate*}   
     The average charge transfer reproduces the mean diffusive current.  

The continuum formulation of the stochastic process is not Lorentz covariant. But, the equations of motion are invariant under Lorentz transformations followed by a reparametrization of the hydrodynamic fields consistent with the derivative expansion.  Indeed, to describe the stochastic dynamics we have adopted a formulation  of  hydrodynamics developed to describe hydrodynamics without boosts~\cite{Novak:2019wqg,deBoer:2020xlc,Armas:2020mpr}. In particular we made considerable use of ``density frame''  of \cite{Armas:2020mpr}.

A notable feature of the density frame is that the charge in a fluid cell ($J^0$ in this case) is sufficient to determine the associated chemical potential, $\mu\equiv J^0/\chi u^0$.  By contrast, in the Landau frame the charge and the three current $J^i$ are both needed, $\mu_\LF =-u_{\mu} J^{\mu}/\chi$. Because of this feature, the equations of motion are first order in time, and do not need any auxiliary variables such  as the diffusive current  or, in full hydrodynamics, the viscous stress tensor  $\pi^{\mu\nu}$. The approach is numerically stable for the diffusion equation and  is expected to be stable for full hydrodynamics, providing a practical way to simulate both stochastic and noise-averaged relativistic fluids in heavy ion collisions.  The equations of motion do not obey Lorentz causality, since the equations do not model the high momentum modes which lie outside of the validity of hydrodynamics. These high momentum modes are essential for Lorentz causality~\cite{Heller:2022ejw,GAVASSINO2023137854,Gavassino:2023mad,Hoult:2023clg,Wang:2023csj}.   In the density frame approach these Fourier modes are simply increasingly damped as $\sim \exp(-D^{ij} k_i k_j t)$ and do not affect the long wavelength dynamics.

In spite of these ``problems'', the density frame dynamics describes well the diffusion of charge in a highly boosted fluid with  $\gamma=10$. Even in the regime where the mean free path becomes comparable to the system size, the leading order density frame predictions remain qualitatively well behaved.  For a specific test case described in Sec.~\ref{sec:comparison_kinetics}, we were able to work out all higher order terms of the density frame gradient expansion, which in the regime of validity of the hydrodynamics, systematically improve the leading order results.   A resummation of this expansion reproduces the causality and Lorentz covariant structure of the dispersion curve of the underlying kinetic model, even though Lorentz causality is violated at any finite order in the gradient expansion.

The next step in this project  is to use the algorithm developed here to
simulate  hydrodynamics in general coordinates and to develop the Metropolis
updates into a practical tool for  stochastic hydrodynamic simulations of heavy
ion collisions.  Indeed, the paradigm of the current paper should work for full
hydrodynamics by implementing the following steps:  \begin{enumerate*}[label=(\roman*)] \item  first take a step  with ideal hydrodynamics; \item  make a proposal to
randomly transfer spatial momentum  between the fluid
cells; \item accept or reject the proposal using the entropy as a weight.\end{enumerate*} In general
coordinates the only complication is that the momentum transfers must be parallel transported from
the cell-faces to the cell-centers before applying the accept-reject criterion.
Recently, many of these steps where used to simulate Model H~\cite{Chattopadhyay:2024jlh}, marking a notable achievement. The study focused on Cartesian incompressible fluids near a critical point with no net momentum, which minimizes the issues related to relativity and covariance. Future work could build on this foundation by developing algorithms for the types of relativistic flows simulated in heavy ion collisions.
It is hoped that the Metropolis algorithm for stochastic hydrodynamics will be
robust and effective, yielding a significant advance in the modeling of the
quark-gluon plasma created in heavy ion collisions. 

\begin{acknowledgments}
G.B. is supported by the National Science Foundation CAREER Award PHY-2143149. 
J.B. and D.T. are supported by the U.S. Department of Energy, Office of Science, Office
of Nuclear Physics, grant No. DE-FG-02-08ER41450.
R.S. acknowledges the support of Polish NAWA Bekker program No. BPN/BEK/2021/1/00342.  
Finally,  we are grateful for the stimulating atmosphere at the ``The Many Faces of Relativistic Fluid Dynamics'' program at the Kavli Institute for Theoretical Physics.
\end{acknowledgments}
\appendix
\section{Advection diffusion in a space-time dependent background flow}
\label{app:spacetimedepend}

  In this appendix  we will consider a charge advecting and diffusing in fluid whose temperature and velocity vary slowly in space and time.   For simplicity we will consider a charge which contributes negligibly to the pressure, e.g. the baryon charge in high energy heavy ion collisions. Thus,  we will drop terms of order the chemical potential squared.

  Before going into details, let us qualitatively describe the result.  First,  we generalize \Sect{subsec:advect_diffuse_density_frame}
  and use the ideal equations of motion to show that the mean diffusive current in the density frame is linearly dependent on the two strains of the system,  $\partial_i \muh$ and $\partial_{(i} \beta_{j)}$. The resulting diffusive current is given in \eqref{eq:Jdfvisc} and  should be compared to  \eqref{eq:JDi} in the body of the text.
  In the Metropolis updates one proposes  correlated charge  \emph{and} momentum transfers between the fluid cells.
  The change in entropy due to the transfers is linearly dependent on  the gradients $\partial_{i} \muh$ and $\partial_{(i} \beta_{j)}$,
  reflecting the fact that $\muh$ and $\beta_i$ are conjugate to charge and momentum.  After applying the  accept-reject 
  criterion of the Metropolis updates, the stochastic dynamics reproduces the mean current of the density frame, \Eq{eq:Jdfvisc}.

  We will now elaborate on this outline, initially ignoring stochastic noise. Following  \Sect{subsec:advect_diffuse_density_frame} we will use ideal equations of motion to eliminate time derivates from from the dissipative current of the Landau Frame.
  The relevant equations of motion follow simply from the conservation of $n/s$ a long the world line of the fluid
  \st
   u^{\mu} \partial_{\mu} (n/s) = 0  \,  ,
  \stp
  which  implies 
  \st
u^{\mu} \partial_{\mu}  \muh    =  \left( \frac{\partial \muh}{\partial \beta } \right)_{n/s} u^{\mu}\partial_\mu \beta  \, .
  \stp
 We next use lowest order equations of motion to  express $u^{\mu}\partial_{\mu} \beta  $ in terms of its spatial strains~\cite{inprep} 
 \st
 u^{\mu} \partial_{\mu}\beta = -u^{\mu} u^{\nu} \partial_{(\mu} \beta_{\nu)}  =   \frac{c_s^2 }{1-c_s^2 v^2 }  (\delta^{ij} - v^i v^j ) \partial_{(i} \beta_{j)} \, ,
 \stp
 which leads to the required approximation 
 \st
 \label{eq:generaldtdmu}
 \partial_t  \muh  \simeq  -v^j \partial_j \muh    +   {\mathfrak c} \left(\delta^{jk} - v^j v^k \right) \partial_{(j}\beta_{k)} \, .
 \stp
 Here 
 \st
 \mathfrak{c} \equiv  \frac{1}{\gamma} \frac{c_s^2 }{1 -c_s^2 v^2 } \left( \frac{\partial \muh}{\partial \beta } \right)_{n/s}   \, ,
 \stp
 is a cross-susceptibility reflecting the coupling between the charge and energy.  
Using   \Sect{subsec:advect_diffuse_density_frame},  we can relate the diffusive current in the density and Landau frames 
\begin{align}
  J^i_{D} =& -T\sigma \left(\Delta^i_\mu- v^i \Delta^0_{\mu} \right)  \Delta^{\mu\nu} \partial_\nu \muh = -T\sigma \left( v^i \partial_t\muh  + \delta^{ij} \, \partial_j \muh \right) \,  , 
\end{align}
and thus with \eqref{eq:generaldtdmu} we find the current in the  density frame\footnote{After noting the world line derivatives
  \[
    \begin{bmatrix}
     -\left( \frac{\partial \beta}{\partial e}\right)_{n/s} &
     \left( \frac{\partial \muh}{\partial e}\right)_{n/s} 
   \end{bmatrix}
   = \frac{\beta}{e + p} 
   \begin{bmatrix}
     \left( \frac{\partial p}{\partial e}\right)_{n} &
     \left( \frac{\partial p}{\partial n}\right)_{e} 
   \end{bmatrix}\, , 
  \]
  and recalling that  $c_s^2 = (\partial p/\partial e)_{n}  + \mathcal O(\muh^2)$, our
\eqref{eq:Jdfvisc} and \eqref{eq:dissipativemat} agree with the eqs.~(90) and (91) from \cite{Armas:2020mpr} in the limit when the 
shear and bulk viscosities are set to zero. }
 \st
 \label{eq:Jdfvisc}
 J^{i}_{D} =  -T\sigma \left[ (\delta^{ij} - v^i v^j) \partial_{j} \muh +  \mathfrak c\,  v^i (\delta^{jk} - v^j v^k )\,  \partial_{(j} \beta_{k)} \right] \, .
 \stp

 Eq.~\ref{eq:Jdfvisc} nicely illustrates expected structure of dissipative stochastic processes. Associated with the diffusive current $J^{i}_D$ and the viscous stress $\Pi^{ij}_D$  are the 
 strains $\partial_i \muh$ and $\partial_{(i} \beta_{j)}$~\cite{Armas:2020mpr,inprep}.  In general there is
 a symmetric matrix of dissipative coefficients connecting the  generalized
 stresses with the strains. In this case the matrix evidently takes the form
 \begin{align}
   \label{eq:dissipativemat}
   \begin{pmatrix}
     J^{i}_D  & \\
    \Pi^{jk}_D 
   \end{pmatrix}
   = &  - 
   \begin{pmatrix}
     T\sigma ( \delta^{i\ell} - v^i v^\ell)  &   T\sigma {\mathfrak c} \, v^i (\delta^{mn} - v^{m} v^{n}) \\
     T\sigma \mathfrak{c}\,  (\delta^{jk} - v^j v^k) v^\ell   &  T\kappa^{jkmn}  
   \end{pmatrix}
   \begin{pmatrix}
     \partial_{\ell} \muh  & \\
     \partial_{(m } \beta_{n)} 
   \end{pmatrix}\, .
 \end{align}
 We have not explicitly worked out the lower left corner of this matrix,  but  have anticipated its form based on the symmetry of the dissipative matrix.  Its contribution to the shear stress is small
\st
\Pi^{jk} \sim  T  \sigma  {\mathfrak c}   \left( \delta^{jk} - v^j v^k \right) v^{\ell}  \partial_{\ell} \muh   \sim O(\muh^2) \,, 
\stp
 and can be neglected  when evaluating the time evolution of the background fluid.
  $T\kappa^{jkmn}$ is proportional to the shear and bulk viscosities of the fluid (without the charge)  and will be reported on separately~\cite{inprep}. This matrix determines the viscous corrections to the background fluid flow, which,  since we are considering a charge moving in a specified background, will be ignored.

 In stochastic fluid dynamics  the dissipative stresses are accompanied by noise
 \begin{align}
   J^i_{D} =&  \bar{J}_{D}^i + \xi^i \, ,   \\
   \Pi^{ij}_{D} =&  \bar{\Pi}_{D}^{ij}  + \xi^{ij} \, ,
\end{align}
with $\bar{J}^i_D$ specified in  \eqref{eq:Jdfvisc}.
The noises are chosen
so that their variance reproduces twice the dissipative  matrix\footnote{There is a general numerical procedure for generating noise with a specified covariance matrix based on the Cholesky decomposition of the matrix~\cite{wiki:Cholesky_decomposition}. }
in \eqref{eq:dissipativemat}.
 The Metropolis algorithm can be used to implement the  dissipative stochastic process.
 We will outline the  necessary steps, limiting the discussion to $1{+}1$ dimensions  where
 the noise matrix takes the simple form 
 \st
 \label{eq:covmatrix1}
  \begin{pmatrix}
    \llangle \xi^x(x) \xi^x(y) \rrangle 
    &
    \llangle \xi^x(x) \xi^{xx}(y) \rrangle  \\

    \llangle \xi^{xx}(x) \xi^x (y)\rrangle 
    &
    \llangle \xi^{xx}(x) \xi^{xx}(y)\rrangle 
  \end{pmatrix}
  = 
  2
   \begin{pmatrix}
     \frac{T\sigma}{\gamma^2 } &   \frac{T\sigma}{\gamma^2} {\mathfrak c}  v^x \\
       \frac{T\sigma}{\gamma^2} \mathfrak{c} v^x   &  T\kappa^{xxxx}  
   \end{pmatrix}
   \delta^2(x -y) \, .
 \stp
 Our goal is to show how the Metropolis algorithm reproduces the mean dissipative current of \eqref{eq:Jdfvisc}.

 The complete algorithm consists of an advective step, discussed in the body of the text, 
 and a  Metropolis step, discussed below.
 The analysis and notation parallels \Sect{sec:num_implement}.
 In the Metropolis step, correlated proposals are  made for the charge and momentum transfers between fluid cells, $Q$  
 and $P$ respectively:
 \st
 Q =  \xi^x \Delta t \,,    \qquad  P = \xi^{xx} \Delta t \, . 
 \stp
 A simple numerical procedure generalizing \Eq{eq:proposal_method} is to take 
 \st
 \xi^x = \sqrt{\frac{2 T \sigma (1-v^2)}{ a_x \Delta t} }  \mathfrak{e}_\parallel\,,   \qquad  \xi^{xx} =\sqrt{ \frac{2 T \sigma(1-v^2)}{a_x \Delta t}} \mathfrak {c}  v^x \mathfrak{e}_\parallel \, , 
 \stp
 where $\mathfrak{e}_\parallel$ is a random number with zero mean variance one. 
 Then in one dimension we loop over the lattice updating pairs of fluid cells, $A$ 
 and $B$ as shown in the figure below:
 \begin{center}
   \includegraphics[width=0.35\textwidth]{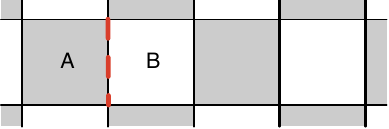}
  \end{center}
  The updates and transfers are 
  \begin{align}
    \N_A \rightarrow&  \N_A + \Delta \N_A \equiv \N_A - Q \, ,  \\
    \N_B \rightarrow&  \N_B + \Delta \N_B \equiv \N_B  + Q \, , \\
    \P_A \rightarrow&  \P_A + \Delta \P_A \equiv \P_A - P \, ,  \\
    \P_B \rightarrow&  \P_B + \Delta \P_B \equiv \P_B +  P  \, .
  \end{align}
 The change in entropy as a result of this  the  proposal is 
 \begin{align}
   \Delta \S =& \sum_{U=A,B} \S(\N_U + \Delta \N_U, \P  + \Delta \P_U) - \S(\N_U, \P_U) \, ,   \\
   \simeq&       -\left[ \partial_x \muh \, \xi^{x} +  \partial_x \beta_x\,  
   \xi^{xx} \right] a_x \Delta t \, .
\end{align}
The proposal is accepted or rejected using the entropy as a statistical weight. 
As in the \Eq{eq:mean_charge_transfer}  the mean charge transfer  takes the form
\begin{align}
  \llangle Q \rrangle_{\rm accept} \simeq &   \llangle Q (1 + \theta ( -\Delta S) \Delta S) \rrangle \, ,   \\
  =&  -\frac{1}{2} a_x \Delta t^2 \, \llangle \xi^x \left(\xi^x \partial_x\muh + \xi^{xx}  \partial_x\beta_x \right) \rrangle \, ,
\end{align}   
which,  
after using the covariance matrix in \eqref{eq:covmatrix1},  reproduces the mean current of the density frame:
\st
\llangle Q  \rrangle_{\rm accept} = \llangle J^x_D \rrangle \Delta t  = - \frac{T\sigma}{\gamma^2} \left[ \partial_x\muh  + \mathfrak{c} v^x \partial_x \beta_x \right]  \Delta t \, .
\stp


\section{Green functions for the kinetic model}
\label{app:green_functions}
In this section we present the Green functions associated with the kinetic model we analyzed in Section \ref{sec:comparison_kinetics}. Let us consider the static case first. As usual, given initial data at some initial time which we set to be $t=0$ the solution to the kinetic equation at a later time $t>0$ is given by propagating the initial data, $(n_0(x^\prime),j_0(x^\prime))$ via the retarded Green function
\begin{equation}
    \left(\begin{matrix}n(t,x) \\ j(t,x)\end{matrix}\right)=\int dx^\prime G_R(t,x-x^\prime)\left(\begin{matrix}n_0(x^\prime) \\ j_0(x^\prime)\end{matrix}\right) \, , 
\end{equation}
where the Green function $G_R$ is a $2\times 2$ matrix which satisfies
\begin{equation}
\left(\begin{matrix} \partial_t && \partial_x \\ \partial_x && 2\lambda + \partial_t \end{matrix}\right)G_R(t-t^\prime,x-x^\prime)=\mathbbm{1}_{2\times 2}\delta(t-t^\prime)\delta(x-x^\prime) \, .
\end{equation}
Here $\lambda=1/(2\tau_R)$ and we set $c_s=1$ for simplicity. By Fourier transforming and with the help of the integral
\begin{equation}
    \int \frac{dk}{2\pi} \frac{\sin\left(\sqrt{k^2-\lambda^2}t\right)}{\sqrt{k^2-\lambda^2}} e^{-i kx}=\frac12\Theta(t^2-x^2) I_0(\lambda \sqrt{t^2-x^2}),
\end{equation}
where $\Theta$ is the Heaviside Theta function and $I_0$ is the modified Bessel function of the first kind, we obtain
\begin{eqnarray}
G_R(t,x)=\frac{\lambda}{2}e^{-\lambda t}\Theta(t^2-x^2)\left(\begin{matrix} G_{nn} && G_{nj} \\ G_{jn} && G_{jj} \end{matrix}\right)+\frac{\lambda}{2}e^{-\lambda t}\left(\begin{matrix} \delta_++\delta_-&& -\delta_++\delta_- \\-\delta_++\delta_- && \delta_++\delta_- \end{matrix}\right)\,.
\end{eqnarray}
Here
\begin{eqnarray}
    \delta_{\pm}&=&\delta(x\pm t),\quad \tau=\sqrt{t^2-x^2} \, ,  \\
    G_{nn}(t,x)&=&\frac\lambda2e^{-\lambda t}\left(\frac t\tau I_1(\lambda \tau)+I_0(\lambda\tau) \right) \, , \\
     G_{nj}(t,x)&=& G_{jn}(t,x)=\frac\lambda2e^{-\lambda t} \left(\frac x\tau I_1(\lambda \tau)\right) \, , \\
      G_{jj}(t,x)&=&\frac\lambda2e^{-\lambda t}\left(\frac t\tau I_1(\lambda \tau)-I_0(\lambda\tau) \right) \, .
\end{eqnarray}
Integrating the singular part of the Green function explicitly we obtain the
\begin{eqnarray}
    n(t,x)&=&\frac{e^{-\lambda t}}{2}\left(n_0(x-t)+n_0(x+t)+j_0(x-t)-j_0(x+t)\right)\nonumber \\
    &&+\int_{x-t}^{x+t} dx^\prime \left( G_{nn}(t,x-x^\prime) n_0(x^\prime)+  G_{nj}(t,x-x^\prime) j_0(x^\prime)\right) \,,
    \\
    j(t,x)&=&\frac{e^{-\lambda t}}{2}\left(n_0(x-t)-n_0(x+t)+j_0(x-t)+j_0(x+t)\right)\nonumber \\
    &&+\int_{x-t}^{x+t} dx^\prime\left( G_{nj}(t,x-x^\prime) n_0(x^\prime)+  G_{jj}(t,x-x^\prime) j_0(x^\prime)\right)\,.
\end{eqnarray}
Since the system is Lorentz covariant, the Green functions for the moving fluid can be obtained by a Lorentz boost. We consider the initial value problem where the initial conditions are still given in the lab frame $t=0$. The charge density and current given in Eq.\eqref{eq:boostedminus} are therefore given by
\begin{eqnarray}
    N(t,x)&=&\frac{e^{-\lambda \kappa_- t}}{2}\left(N_0(x-t)+J_0(x-t)\right)+\frac{e^{-\lambda \kappa_+ t}}{2}\left(N_0(x+t)-J_0(x+t)\right)\nonumber \\
    &&+\gamma \int_{x-t}^{x+t} dx^\prime\left[ \left(G_{nn}(\tilde t-\gamma v x,\gamma x- \tilde x^\prime)+v G_{nj}(\tilde t-\gamma v x,\gamma x- \tilde x^\prime)\right) N_0(x^\prime) \right.
    \nonumber \\
  && \quad \quad \quad \; \left. +  \left(G_{nj}(\tilde t-\gamma v x,\gamma x- \tilde x^\prime)+v G_{jj}(\tilde t-\gamma v x,\gamma x- \tilde x^\prime)\right) J_0(x^\prime)\right] \, ,
    \\
    J(t,x)&=&\frac{e^{-\lambda \kappa_- t}}{2}\left(N_0(x-t)+J_0(x-t)\right)-\frac{e^{-\lambda \kappa_+ t}}{2}\left(N_0(x+t)-J_0(x+t)\right)\nonumber \\
    &&+\gamma \int_{x-t}^{x+t} dx^\prime\left[ \left(G_{nj}(\tilde t-\gamma v x,\gamma x- \tilde x^\prime)+v G_{nn}(\tilde t-\gamma v x,\gamma x- \tilde x^\prime)\right) N_0(x^\prime) \right. \nonumber\\
     &&\quad \quad \quad  \; \left. +  \left(G_{jj}(\tilde t-\gamma v x,\gamma x- \tilde x^\prime)+v G_{nj}(\tilde t-\gamma v x,\gamma x- \tilde x^\prime)\right) J_0(x^\prime)\right] \, .
\end{eqnarray}
where $\kappa_\pm=\sqrt{(1\pm v)/(1\mp v)}$, $\tilde t= \gamma (t + v x^\prime)$ and $\tilde x^\prime=\gamma (x^\prime + v t)$. 
\endgroup

\end{thebibliography}
\end{document}